\documentclass[
prl
,twocolumn
,nofootinbib
,floatfix
,superscriptaddress
]{revtex4-2} 
\usepackage{notes2bib}
\usepackage{natbib}
\usepackage{gensymb}
\usepackage{amsthm}
\usepackage{amsmath}
\usepackage{color}
\usepackage{bm}
\usepackage{amsmath,amssymb}
\usepackage{amsfonts}
\usepackage{dsfont}
\usepackage{graphicx}
\usepackage{float}
\usepackage{subfigure}
\usepackage{verbatim}
\usepackage{amsfonts}
\usepackage{bbold}
\usepackage{dcolumn}
\usepackage{bm}
\usepackage[dvipsnames]{xcolor}
\usepackage{listings}
\definecolor{blueprl}{RGB}{46,48,146}
\usepackage[pdftex,colorlinks=true,urlcolor=blueprl,citecolor=blueprl,linkcolor=blueprl]{hyperref}
\usepackage{mathrsfs}
\usepackage{mathtools}
\usepackage{times}
\usepackage{color}
\usepackage[dvipsnames]{xcolor}

\usepackage{braket}

\usepackage{soul}

\newcommand{\mtx}{\textsf{x}}
\newcommand{\dd}{\dagger}

\usepackage{mathtools}
\usepackage{dsfont}
\usepackage[mathscr]{euscript}
\usepackage{amsmath}
\usepackage{graphicx}
\usepackage{dcolumn}
\usepackage{bm}
\usepackage{epsfig}
\usepackage{amssymb,latexsym,mathrsfs}
\usepackage{graphicx}
\usepackage{color}
\usepackage{hyperref}
\usepackage{float}
\usepackage{diagbox}
  \usepackage{cancel}
  \usepackage[normalem]{ulem}

\definecolor{vividviolet}{rgb}{0.62, 0.0, 1.0}
\definecolor{amaranth}{rgb}{0.9, 0.17, 0.31}
\definecolor{palatinateblue}{rgb}{0.15, 0.23, 0.89}
\definecolor{brightpink}{rgb}{1.0, 0.0, 0.5}
\definecolor{cornflowerblue}{rgb}{0.39, 0.58, 0.93}
\definecolor{deepcarminepink}{rgb}{0.94, 0.19, 0.22}
\definecolor{radicalred}{rgb}{1.0, 0.21, 0.37}
\definecolor{blueblue}{RGB}{21,47,181}
\definecolor{greengreen}{RGB}{65,166,16}

\usepackage{xtab,afterpage}
\usepackage{hyperref}
\usepackage{setspace,calc}

\allowdisplaybreaks

\hypersetup{ linktoc=all,
    colorlinks, linkcolor={blueblue},
    citecolor={greengreen}, urlcolor={blueblue}
}

\newcommand{\be}{\begin{equation}}
\newcommand{\ee}{\end{equation}}
\newcommand{\bs}{\begin{split}} 
\newcommand{\bea}{\begin{eqnarray}}
\newcommand{\eea}{\end{eqnarray}}
\newcommand{\infint}{\int_{-\infty }^{\infty }}
\newcommand{\vt}{\vphantom{\frac{1}{2}}}

\newcommand{\non}{\nonumber} 
 
\newcommand{\p}{\partial} 
\newcommand{\D}{\mathrm{d}}
\newsavebox{\myhbar}

\begin{document}

\title{Quantum superpositions of Minkowski spacetime}
\author{Joshua Foo}
\email{joshua.foo@uqconnect.edu.au}
\affiliation{Centre for Quantum Computation \& Communication Technology, School of Mathematics \& Physics, The University of Queensland, St.~Lucia, Queensland, 4072, Australia}
\author{Cemile Senem Arabaci}
\affiliation{Department of Physics and Astronomy, University of Waterloo, Waterloo, Ontario, Canada, N2L 3G1}
\author{Magdalena Zych}
\affiliation{Centre for Engineered Quantum Systems, School of Mathematics and Physics, The University of Queensland, St. Lucia, Queensland, 4072, Australia}
\author{Robert B.\ Mann}
\affiliation{Department of Physics and Astronomy, University of Waterloo, Waterloo, Ontario, Canada, N2L 3G1}
\affiliation{Perimeter Institute, 31 Caroline St., Waterloo, Ontario, N2L 2Y5, Canada}

\begin{abstract}
Within any anticipated unifying theory of quantum gravity, it should be meaningful to combine the fundamental notions of quantum superposition and spacetime to obtain so-called ``spacetime superpositions'': that is, quantum superpositions of different spacetimes not related by a global coordinate transformation. Here we consider the quantum-gravitational effects produced by superpositions of periodically identified Minkowski spacetime (i.e.\ Minkowski spacetime with a periodic boundary condition) with different characteristic lengths. By coupling relativistic quantum matter to fields on such a spacetime background (which we model using the Unruh-deWitt particle detector model), we are able to show how one can in-principle ``measure'' the field-theoretic effects produced by such a spacetime. We show that the detector's response exhibits discontinuous resonances at rational ratios of the superposed periodic length scale. 
\end{abstract} 

\date{\today} 

\maketitle 

\textit{Introduction}\textemdash In the absence of a full-fledged theory of quantum gravity, there has been increasing interest in studying the phenomenology of quantum gravity using operational approaches. An operational approach is one that grounds such phenomena in measurements of physical observables using tools such as detectors, rods, and clocks. Some recent investigations in the field of relativistic quantum information (RQI) and quantum field theory in curved space (CS--QFT) have combined fundamental features of quantum theory, such as the notions of superposition, entanglement, measurement, with those of general relativity, such as proper time, causal structure, and spacetime, to study physical effects that would be otherwise out of reach with current top-down approaches such as string theory \cite{gubser1998gauge,seiberg1999string,polchinski1998string,witten1995string} and loop quantum gravity \cite{rovelli2008loop,rovelli2014covariant,thiemann2003lectures}. Recent investigations include those which explore the quantization of time using a ``clock'' moving in a superposition of localized momenta \cite{smith2020quantum}, the reconstruction of the spacetime metric in terms of quantum field correlations \cite{10.3389/fphy.2021.655857}, and the violation of classical constraints on causal order due to superpositions of massive bodies \cite{zych2019bell}. Rather than pursuing a complete theory from the top-down, these investigations exemplify ``bottom-up'' approaches for studying quantum-gravitational physics.  

In this paper, we adopt this perspective  in order to study an important problem in quantum gravity, namely quantum superpositions of spacetime. Assuming such a theory exists, we expect such superpositions of ``semiclassical spacetime states'' (i.e.\ each amplitude of the superposition corresponding to a classical matter configuration associated with a classical manifold and gravitational field) to be valid solutions \cite{christodoulou2019possibility,belenchiaPhysRevD.98.126009,Giacomini2021spacetimequantum,giacomini2022quantum}. More specifically, we are interested in the kinds of superpositions in which the respective amplitudes are not related by a global coordinate transformation and are hence diffeomorphic. We have recently argued that the environment generated by such ``spacetime superpositions'' does not meaningfully differ from those generated on a single ``classical'' background in which quantum systems residing within are prepared and measured in appropriate quantum states \cite{Foo_2021,foo_2022}. 

Instead, we are primarily interested in superpositions of spacetimes that are not diffeomorphic invariant i.e.\ the individual amplitudes represent unique solutions to Einstein's field equations. As explained, we do not propose a full quantum-gravitational theory for the emergence of such superpositions, but assume that they are valid solutions within an anticipated theory. Our goal then is to study the operational effects induced upon quantum matter residing within such spacetimes. In a recent paper, we studied the quantum-gravitational effects produced by a Banados-Teitelboim-Zanelli (BTZ) black hole in a superposition of masses and showed that a quantum detector, modelled as a qubit linearly coupled to a massless scalar field, was sensitive to the mass ratio of the superposed spacetime amplitudes \cite{foo_2022}. Intriguingly, such ratios were found to be commensurate with those allowed by Bekenstein's conjecture for the horizon area quantization of black holes in quantum gravity, suggesting that such detectors are sensitive to the signatures of quantum-gravitational effects \cite{bekenstein2020quantum,bekensteinPhysRevD.7.2333}. 

Building off this result, we apply our method to study the operational effects produced by quantum superpositions of Minkowski spacetime. This setting is perhaps the simplest in which the physical effects of spacetime superpositions can be studied, and so offers considerable insight into this phenomenon.
More specifically, we consider (3+1)-dimensional Minkowski spacetime with a periodic boundary condition imposed along one spatial dimension. We then consider a scenario where the characteristic length scale of this periodicity is in a quantum-controlled superposition of lengths. Constructing a quantum field theory on such a spacetime is similar to the procedure applied to the BTZ black hole, which is constructed from periodic identifications of anti-de Sitter space.
We thus expect that similar results found in the black hole superposition should appear in this setting.

Our aims   are three-fold. First, we provide a new example of our framework for analyzing the effects induced by spacetime superpositions upon quantum matter in its simplest setting, namely superpositions of topologically identified Minkowski space. Indeed, we show how this scenario produces effects related to those produced by the mass-superposed BTZ black hole. Second, we address some technical issues regarding how one should go about performing quantum field theory calculations in settings involving superpositions of spacetime. Specifically, the conceptually simple example of superposed Minkowski spacetime allows us to clarify the choice of vacuum state when calculating correlation functions for fields quantized on this superposed background. Finally, we propose a toy-model for a particle detector residing in a superposition of spacetimes, drawing an analogy between the periodically identified Minkowski spacetime { and} %with 
a quantum field with periodic boundary conditions {(%e.g.\ modelled by the nontrivial topology of a
which could be realisable using a toroidal cavity in optomechanical setups \cite{bowen2015quantum})}. %The superposition of topological identifications of the spacetime is thus analogous to the periodic field whose boundary conditions are imposed in a superposition of characteristic lengths. %While there are points at which the analogy breaks down, 
 Further study of such setups may open a promising route  %may motivate further discussion concerning the 
towards simulations of the effects produced by quantum superpositions of spacetimes on quantum fields, which is a topic of growing interest~%of which examples utilizing Bose-Einstein condensates have recently been proposed 
\cite{barcelo2021superposing}.

Our paper is organized as follows. We first review the theory of quotient spaces in Minkowski spacetime, and the construction of automorphic fields on these spaces. We then review the model for coupling a UdW detector to a quantum-controlled superposition of spacetimes, and apply this to the topologically identified Minkowski spacetime. Next, we present results concerning the detector's response to a massless scalar field in the superposed Minkowski spacetime. We introduce our toy-model for a detector coupled to a quantum field with a superposed periodic boundary condition imposed. We conclude with some final thoughts. Throughout this article, we utilise natural units, $\hslash = k_B = c = G = 1$.  

\textit{Quotient Spaces of Minkowski spacetime}\textemdash \label{sec:II}
In this section we review the basic geometric elements of Minkowski spacetime $M$ and its periodically identified quotient space $M_0$, before introducing the quantization scheme of the automorphic fields used to calculate two-point correlation functions and detector transition probabilities. 

Let us begin with the familiar (3+1)-dimensional Minkowski spacetime $M$, parametrized by the usual coordinates ($t,x,y,z$) with line element 
\begin{align}
    \D s^2 &= \D t^2  - \D x^2 - \D y^2 - \D z^2,
\end{align}
where the metric signature is chosen to be $(+,-,-,-)$ for straightforward comparison with existing literature \cite{martinmartinezPhysRevD.93.044001}. We consider a massless scalar field $\hat{\phi}$ that is a solution to the Klein-Gordon equation $\Box \hat{\phi}(\mtx) = 0$ and may be expanded in the plane wave basis 
\begin{align}
    \hat{\phi}(\mtx) &= \int\frac{\D k^3}{(2\pi)^{3/2}}\frac{1}{\sqrt{2|\textbf{k}|}} \left( e^{-i|\textbf{k}|t + i \textbf{k}\cdot \textbf{x}} \hat{a}_\textbf{k} + \mathrm{H.c} \right)
\end{align}
where $\textbf{k} = (k_x, k_y, k_z)$, $\textbf{x} = (x,y,z)$ are the momentum and position three-vectors respectively, and $\hat{a}_\textbf{k}(\hat{a}_\textbf{k}^\dd)$ are annihilation (creation) operators of a single-frequency mode. Letting $| 0 \rangle$ denote the Minkowski vacuum state annihilated by $\hat{a}_\textbf{k}$, it can be shown that the two-point correlation function
\begin{align}
    W_M(\mtx,\mtx') \equiv \langle 0 | \hat{\phi}(\mtx) \hat{\phi}(\mtx') | 0 \rangle , \vt 
\end{align}
pulled back to the worldlines $\mtx$, $\mtx'$ is given by \cite{martinmartinezPhysRevD.93.044001}
\begin{align}\label{3}
    W_M(\mtx, \mtx' ) &= \frac{1}{4\pi i }\mathrm{sgn}(t-t') \delta ( \sigma(\mtx, \mtx') ) - \frac{1}{4\pi^2\sigma(\mtx, \mtx') }
\end{align}
where $W_M(\mtx, \mtx') := \langle 0 | \hat{\phi}(\mtx) \hat{\phi}(\mtx' ) | 0 \rangle$, $\mathrm{sgn}(t-t')=\pm 1$ depending on the sign of $t-t'$, and the geodesic distance $\sigma(\mtx, \mtx')$ on single spacetime is given by 
\begin{align}
    \sigma(\mtx, \mtx') &= (t-t')^2 -(x-x')^2 -(y-y')^2 - (z-z')^2. 
\end{align}
Equation (\ref{3}) is commonly referred to as the Wightman function, pulled back to the worldlines $(\mtx, \mtx')$. 

\textit{The $M/J_0$ Quotient Space}\textemdash The flat spacetime $M_0  = M /J_0$  is built as a quotient of $M$ under the isometry group $Z \simeq {J_0^n}$, $J_0: (t,x,y,z) \mapsto (t,x,y,z+l)$ \cite{Banach:1979iy,Banach_1980}. Henceforth we refer to $M_0$ as a cylindrical spacetime with circumference $l$. $J_0$ preserves space and time orientation and acts freely and properly, ensuring that $M_0$ is a space and time orientable Lorentzian manifold \cite{martinmartinezPhysRevD.93.044001}. To construct a quantum field theory on the quotient space, we define the automorphic field $\hat{\psi}(\mtx)$ constructed from the usual massless scalar field $\hat{\phi}(\mtx)$ as the image sum \cite{langlois2006causal}
\begin{align}
    \hat{\psi}(\mtx) &= \frac{1}{\sqrt{\mathcal{N}}} \sum_n \eta^n \hat{\phi}(J_0^n \mtx) 
\end{align}
where $\mathcal{N} = \sum_n \eta^{2n}$ is a normalisation factor that ensures that 
\begin{align}
    \left[ \hat{\psi}(\mtx), \hat{\dot{\psi}}(\mtx') \right] &= \delta(\mtx- \mtx') + \mathrm{image\:terms},
\end{align}
and $\eta = \pm 1$ denotes an untwisted (twisted) field. To obtain the Wightman functions, we have
\begin{align}
    W_{J_0}^{(D)} (\mtx, \mtx') &= \frac{1}{\mathcal{N}} \sum_{n,m} \eta^n \eta^m  W_M (J_{0_D}^n \mtx, J_{0_D}^m \mtx' ),
    \nonumber \\
    &= \frac{1}{\mathcal{N}} \sum_{n,m} \eta^n ( \eta^n \eta^m ) W_M (J_{0_D}^n \mtx, J_{0_D}^n J_{0_D}^m \mtx'), \nonumber 
    \\
    &=\frac{1}{\mathcal{N}} \sum_{n,m} \eta^{2n} \eta^m W_M (\mtx, J_{0_D}^m \mtx'),
    \nonumber 
    \\ \label{9}
    &= \sum_m \eta^m W_M (\mtx, J_{0_D}^m \mtx') \vphantom{\frac{1}{\sum_n\eta^{2n}}},
\end{align}
where our superscript notation $D = A,B$ anticipates our eventual goal of computing functions associated with the field quantized on $M_0$ with two characteristic lengths $l_A$ and $l_B$ respectively. Specifically, $J_{0_A}^n$ and $J_{0_B}^m$ denote the respective isometries 
\begin{align}
    J_{0_A}^n &: (t,x,y,z) \mapsto (t,x,y,z + l_A) ,
    \vt 
    \\
    J_{0_B}^m &: (t,x,y,z) \mapsto (t,x,y,z+l_B) .
    \vt 
\end{align}
It is important to note that the evaluation of Eq.\ (\ref{9}) occurs with respect to the Minkowski vacuum state. The identification of the spacetime enforcing periodicity in the $z$-direction can be understood as the action of the operator $J_{0_D}^n$ %acting 
on the coordinates of the field. Furthermore, while it is common to use the simplified form of $W_{J_0}$ shown in (\ref{9}), such a treatment is inadequate when considering superpositions of spacetime. That is, for superpositions of the characteristic length of the quotient space $M/J_0$, one must construct correlation functions that %are essentially
arise from superpositions of the different topological identifications, which generate two \textit{different} discrete isometries on the field. %in superposition. 
A recent investigation \cite{kabel} considers a related question of the quantization of scalar Klein-Gordon field on a superposition of non-diffeomorphic backgrounds from the perspective of quantum reference frame transformations. 

For quantum-controlled superpositions of two cylindrical spacetimes, the resulting amplitudes contain also Wightman functions given by:
\begin{align}
    W_{J_0}^{(AB)} (\mtx_A, \mtx_B') &= \frac{1}{\mathcal{N}} \sum_{n,m} \eta^n \eta^m W_M ( J_{0_A}^n \mtx, J_{0_B}^m \mtx') 
\end{align}
where  
\begin{align}
    W_M(J_{0_A}^n \mtx, J_{0_B}^m \mtx' ) &= \langle 0| \hat{\phi}(J_{0_A}^n \mtx) \hat{\phi}(J_{0_B}^m \mtx' ) | 0 \rangle 
\end{align}
is evaluated with respect to a single vacuum state, $| 0 \rangle$. While one could conceive of a scenario in quantum gravity where the vacuum state itself is quantum-controlled, this simple case does not require such an assumption.  
The effects arising from superposed quantum amplitudes of the spacetime here occur solely through the action of the two different discrete isometries $J_{0_A}^n$, $J_{0_B}^m$. We recently made a similar assumption in utilizing the ``global'' ground state of the field in anti-de Sitter space to evaluate correlation functions in a BTZ spacetime in a superposition of masses~\cite{foo_2022}. 

Returning to the Wightman functions, we have explicitly that 
\begin{align}\label{11}
    W_{J_0}^{(D)} (\mtx, \mtx') &= \frac{1}{\mathcal{N}} \sum_{n,m} \eta^n \eta^m \bigg[ \frac{\mathrm{sgn}(t-t') \delta( \sigma(J_{0_D}^n \mtx, J_{0_D}^m\mtx' ) )}{4\pi i } \non \\
    &\qquad - \frac{1}{4\pi^2 \sigma( J_{0_D}^n \mtx, J_{0_D}^m \mtx' ) } \bigg] \\ \label{12}
    W_{J_0}^{(AB)} (\mtx, \mtx' ) &= \frac{1}{\mathcal{N}} \sum_{n,m} \eta^n \eta^m \bigg[ \frac{\mathrm{sgn}(t-t') \delta( \sigma( J_{0_A}^n \mtx, J_{0_B}^m \mtx') )}{4\pi i } \non \\
    &\qquad - \frac{1}{4\pi^2 \sigma ( J_{0_A}^n \mtx, J_{0_B}^m \mtx' ) } \bigg] 
    % \\ \label{13}
    % W_{J_0}^{(AA)} (\mtx, \mtx' ) &= \frac{1}{\mathcal{N}} \sum_{n,m} \eta^n \eta^m \bigg[ \frac{\mathrm{sgn}(t-t') \delta( \sigma( J_{0_A}^n \mtx, J_{0_A}^m \mtx') )}{4\pi i } \non \\
    % & - \frac{1}{4\pi^2 \sigma ( J_{0_A}^n \mtx, J_{0_A}^m \mtx' ) } \bigg] \\ \label{14}
    %  W_{J_0}^{(BB)} (\mtx, \mtx' ) &= \frac{1}{\mathcal{N}} \sum_{n,m} \eta^n \eta^m \bigg[ \frac{\mathrm{sgn}(t-t') \delta( \sigma( J_{0_B}^n \mtx, J_{0_B}^m \mtx') )}{4\pi i } \non \\
    % & - \frac{1}{4\pi^2 \sigma ( J_{0_B}^n \mtx, J_{0_B}^m \mtx' ) } \bigg] 
\end{align}
where the geodesic distances are respectively
\begin{align}
    \sigma(J_{0_D}^n x, J_{0_D}^m x' ) &= (t-t')^2 - l_D^2(n-m)^2, \vt \\
    \sigma( J_{0_A}^n x , J_{0_B}^m x' ) &= (t-t')^2 - (l_An - l_Bm)^2, \vt
    % \\
    % \sigma( J_{0_A}^n x , J_{0_A}^m x' ) &= (t-t')^2 - (l_An - l_Am)^2 \vt \\
    % \sigma( J_{0_B}^n x , J_{0_B}^m x' ) &= (t-t')^2 - (l_Bn - l_Bm)^2 \vt
\end{align}
and we have considered, without loss of generality, a detector static at the origin of the coordinate system of both spacetimes. 

\textit{Unruh-deWitt detector in superposed Minkowski spacetime}\textemdash 
We are interested in studying the effects induced by the superposed Minkowski spacetime upon relativistic quantum matter which is coupled to the spacetime through its interaction with a (massless scalar) quantum field. We can describe this system in the Hilbert space $\mathcal{H} = \mathcal{H}_S \otimes \mathcal{H}_F \otimes \mathcal{H}_M$ which is a tensor product of the spacetime, quantum field, and matter degrees of freedom respectively. 

For simplicity, let us consider the topologically identified $M_0$ spacetime in a superposition of two characteristic lengths $l_A$ and $l_B$, and the field in the Minkowski vacuum state. We likewise introduce a simple particle detector model (the aforementioned Unruh-deWitt model) to describe our quantum matter coupled to the field and spacetime. The point-like, two-level detector is assumed to be initially in its ground state $| g \rangle$, such that the initial state of the combined system is given by 
\begin{align}
    | \psi(t_i) \rangle &= \frac{1}{\sqrt{2}} ( |l_A \rangle + |l_B \rangle ) | 0 \rangle |g \rangle 
\end{align}
The coupling between the spacetime superposition, field, and detector is described by the following interaction Hamiltonian \cite{fooudw1PhysRevD.102.085013,fooudw2PhysRevResearch.3.043056,fooudw3PhysRevD.103.065013,fooudw5https://doi.org/10.48550/arxiv.2204.00384}:
\begin{align}
    \hat{H}_\mathrm{int.} &= \lambda \eta(\tau) \hat{\sigma}(\tau ) \sum_{D=A,B} \hat{\psi}(\mtx_D ) \otimes | l_i \rangle\langle l_i | .
\end{align}
Here, $\lambda\ll 1$ is a coupling constant, $\tau$ is the proper time in the detector's reference frame, $\eta(\tau)$ a time-dependent switching function that mediates the interaction, $$\hat{\sigma}(\tau) = |e \rangle\langle g| e^{i\Omega \tau} +
|g \rangle\langle e| e^{-i\Omega \tau}
$$ 
is the SU(2) ladder operator between the detector's internal states $|g\rangle$, $|e\rangle$ with energy gap $\Omega$, and $\hat{\phi}(\mtx_i)$ is the field operator pulled back to the worldline $\mtx$ parametrised by the coordinates of the detector and the topology of the spacetime. The projector $| l_i \rangle \langle l_i |$ acts as a quantum control for the spacetime. This could be some ancillary system which is entangled with the spacetime, and can be ideally time-evolved and measured in a Mach-Zehnder-type interferometer. For simplicity, we need not posit such an ancilla, and assume, as other recent studies have, that a ``measurement'' can be performed that allows one to witness interference effects between the spacetime amplitudes in superposition \cite{howlhttps://doi.org/10.48550/arxiv.2203.05861}.

% \begin{figure}[h]
%     \centering
%     \includegraphics[width=0.65\columnwidth]{Fig1.png}
%     \caption{Visual representation of the spacetime superposition of interest. The Unruh-deWitt detector (depicted as a two-level atom) is situated at the coordinate origin of both spacetimes. The `spacetime amplitudes' of the superposition have different characteristic lengths, $l_A$ and $l_B$. }
%     \label{fig:my_label}
% \end{figure}

Formally, the basis states $|l_i\rangle$ are energy eigenstates of the free Hamiltonian where $\hat{H}_{0,S} | l_i \rangle = E_i | l_i \rangle$ where $E_i$ are the energies associated with the periodic length $l_i$. This will generally introduce a time-dependent phase to the evolution of the superposition. For simplicity, it is instructive to consider a rotating frame transformation \cite{whaleyPhysRevA.29.1188} for which the evolution of the superposition state is ``frozen'' to the initial phase relationship. Such an assumption greatly simplifies the calculations without losing a significant amount of insight into the problem. The time-evolution operator, 
\begin{align}
    \hat{U} &= \hat{\mathcal{T}} \exp \left( - i \infint \D \tau \: \hat{H}_\mathrm{int.}(\tau) \right) 
\end{align}
can be expanded perturbatively in the Dyson series as follows:
\begin{align}
    \hat{U} &= \mathds{I} - i \lambda \infint\D \tau \: \hat{H}_\mathrm{int.} 
    \non 
    \\
    & - \lambda^2 \infint\D \tau \int_{-\infty}^\tau \D \tau' \hat{H}_\mathrm{int.}(\tau) \hat{H}_\mathrm{int.}(\tau') + \mathcal{O}(\lambda^3)
\end{align}
We evolve the initial state in time,
\begin{align}
    \hat{U} | \psi(t_i) \rangle &= \frac{1}{\sqrt{2}} ( \hat{U}_A | l_A \rangle + \hat{U}_B | l_B \rangle ) | 0 \rangle |g \rangle 
\end{align}
before measuring the control state in the superposition basis $(|l_A \rangle \pm |l_B \rangle)/\sqrt{2}$ and tracing out the final field states. This leaves the following result for the %joint transition probability of 
final state of the detector, 
\begin{align}\label{eq23}
    \hat{\rho}_D &= \begin{pmatrix}
    1 - P_G^{(\pm)} & 0 \\ 0 & P_E^{(\pm)} 
    \end{pmatrix}.
\end{align}
Note that the state   \eqref{eq23} is not normalized, since we are considering final \textit{conditional} state of the detector. The transition probability of the detector is more specifically given by 
\begin{align}
    P_E^{(\pm)} &= \frac{\lambda^2}{4} \Big( P_A + P_B \pm 2L_{AB} \Big) 
\end{align}
where 
\begin{align}\label{25}
    P_D &= \infint\D \tau \infint\D \tau' \chi(\tau) \overline{\chi}(\tau') W_{J_0}^D(\mtx, \mtx') 
\end{align}
is the transition probability of a single detector in a cylindrical spacetime with characteristic length $l_D$ $(D = A,B)$, and 
\begin{align}\label{26}
    L_{AB} &= \infint\D \tau \infint\D \tau' \chi(\tau) \overline{\chi}(\tau') W_{J_0}^{AB} (\mtx, \mtx' ) 
\end{align}
is a cross-correlation term between the field on the background spacetime in a superposition of two characteristic lengths, $l_A$ and $l_B$. We have also defined 
\begin{align}
    \chi(\tau) &= \exp \left( - \frac{\tau^2}{2\sigma^2} \right) e^{-i\Omega\tau} 
\end{align}
as the Gaussian switching function with characteristic width $\sigma$. The introduction of a time-dependent switching function is necessary for a particle detector in flat Minkowski spacetime to detect any field quanta, since a detector that is eternally interacting with the field will remain in its ground state. The result is sometimes interpreted as a manifestation of the energy-time uncertainty principle, in which rapidly switched interactions may promote virtual vacuum fluctations into the detection of real field quanta (thus exciting the detector) \cite{menicucciPhysRevD.79.044027}. Finally, it is important to note that if one traces out the control rather than measuring it in a superposition basis, the detector transition probability becomes a classical mixture of the individual contributions from spacetime amplitude $A$ and $B$:
\begin{align}
    P_E^\mathrm{(Tr)} &= \frac{\lambda^2}{2} \Big( P_A + P_B \Big) .
\end{align}
Returning to the conditional transition probability (given the control is measured in $|\pm \rangle$), we can insert the Wightman functions, Eq.\ (\ref{11}) and (\ref{12}), into Eq.\ (\ref{25}) and (\ref{26}), to obtain the ``local'' contribution to the transition probability, given by  
\begin{align}\label{27}
    P_D &= P_M + \frac{\sigma}{4\sqrt{\pi}l_D \sum_n \eta^{2n}} \Big[ S_1 - S_2 \Big] 
\end{align}
where 
\begin{align}
    S_1 &= \sum_{n\neq m } \frac{e^{-\frac{l_D^2(n-m)^2}{4\sigma^2}}}{n-m} \mathrm{Im} \left[ e^{il_D(n-m)\Omega} \mathrm{erf} \left( \frac{il_D(n-m)}{2\sigma } + \sigma\Omega \right) \right] \nonumber 
    \\
    S_2 &= 2 \sum_{n>m} \frac{e^{-\frac{l_D^2(n-m)^2}{4\sigma^2}}}{n-m} \sin( \Omega l_D (n-m) ) \bigg] \nonumber
\end{align}
and
\begin{align}
    P_M &= \frac{1}{4\pi} \bigg[ e^{-\sigma^2\Omega^2} - \sqrt{\pi}\sigma\Omega \mathrm{erfc} ( \sigma\Omega ) \bigg]
\end{align}
is the transition probability of a single detector in flat Minkowski spacetime with no identifications. Equation (\ref{27}) is equivalent to the expression studied in \cite{PhysRevD.93.044001} for the single-detector transition probability in the $M_0$ spacetime. We see that there is a  Minkowski contribution and an image sum contribution that  accounts for the possible identifications in the $M/J_0$ space. 

Meanwhile, the cross-correlation term is given by 
\begin{align}
    \label{30}
    L_{AB} &= \frac{K_\gamma}{\sum_n \eta^{2n}} P_M + \frac{\sigma}{4\sqrt{\pi} \sum_n \eta^{2n}} \Big[ J_1 - J_2 \Big]  
\end{align}
where
\begin{align}
    J_1 &= \sum_{l_{nm} \neq 0} \frac{e^{- \frac{l_{nm}^2}{4\sigma^2}}}{l_{nm}} \mathrm{Im} \bigg[ e^{i l_{nm} \Omega}\; \mathrm{erf} \left( \frac{il_{nm}}{2\sigma}  + \sigma\Omega \right) \bigg]
    \\
    J_2 &= 2 \sum_{l_{nm}>0} \frac{e^{-\frac{l_{nm}^2}{4\sigma^2}}}{l_{nm}} \sin(\Omega l_{nm})
\end{align}
and $l_{nm} = l_A n - l_B m$. We have also defined 
\begin{align}
   K_\gamma = \mathsf{coeff} \left( \sum_{n,m} f\left( n - \gamma m \right) , f(0) \right) 
\end{align}
where $\gamma = l_B/l_A$ is the ratio of the cylindrical spaces in superposition, and $\mathsf{coeff}(x(y),y)$ is the coefficient of $y$ in the function $x(y)$. The appearance of this function results from the evaluation of the image sum contributions to $L_{AB}$. Notice in particular that the Wightman function for the cross-term, Eq.\ (\ref{12}), contains multiple singular points whenever $n  - \gamma m  = 0$. These poles are treated differently in comparison to the single pole  in the Minkowski Wightman function (see Appendix). Thus, when summing over the identification variables $(n,m)$, one ``pulls out'' a Minkowski contribution to the total expression for $L_{AB}$, whenever $n - \gamma m = 0$. The appearance of these ``resonances'' whenever $n = \gamma m$ is similar to the result previously derived for the BTZ spacetime. In that scenario, the mass ratios for which a ``resonance'' appeared were commensurate with those predicted by Bekenstein in his famous quantum black hole conjecture, wherein the black hole mass is treated as a quantum number. 

\textit{Results}\textemdash We are now able to plot the response of the detector to the field, situated in this universe in a superposition of topologies. 

\begin{figure}[h]
    \centering
    \includegraphics[width=0.9\linewidth]{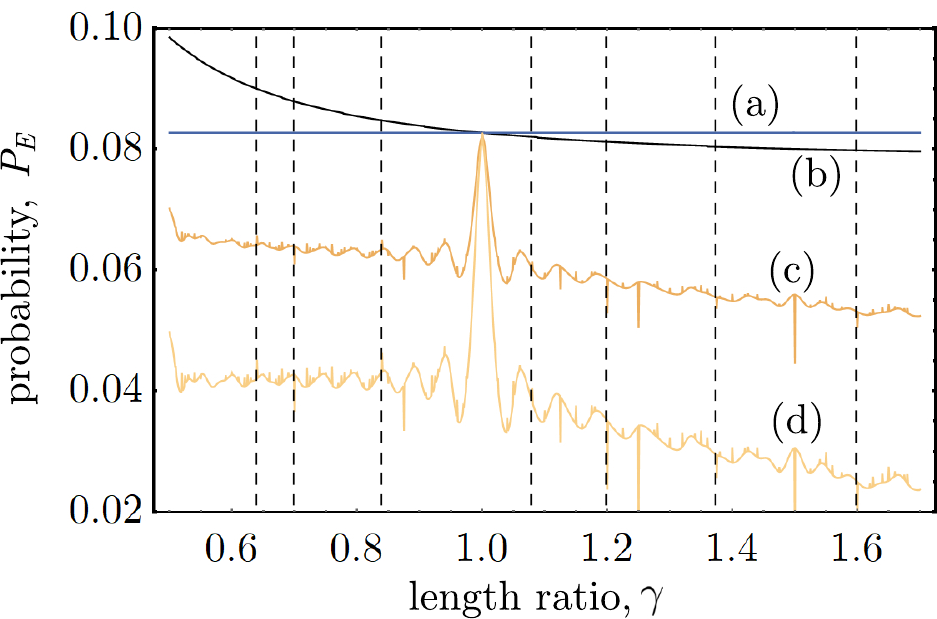}
    \caption{Contributions to the total probability, (c), of the detector after the control is measured in the $| + \rangle$ state, as a function of $\gamma$. We have marked out a few values at which resonances in the transition probability are visible. (a) corresponds to the contribution from one of the spacetime amplitudes (with fixed length $l_A = 1$), (b) corresponds to the contribution from the other amplitude whose length $l_B$ varies, while (d) is the interference term between the two spacetime amplitudes. We have also chosen $\Omega \sigma = 1/100$.}
    \label{fig:PE}
\end{figure}

In Fig.\ \ref{fig:PE}, we have plotted the transition probability of the detector as a function of $\gamma$, the ratio of the characteristic lengths of the superposed spacetimes. There are several physical features of the transition probability worth noting. Most interestingly, we observe discontinuous resonant peaks in the transition probability at rational values of $\gamma$, where some of these values are marked with a vertical line in Fig.\ \ref{fig:PE}. In reality, we expect a countably infinite number of these discrete peaks at every rational value of $\gamma$, just by inspecting the discontinuous form of the interference term in Eq.\ (\ref{30}). The magnitude of these peaks may not necessarily be visible; moreover, we are limited by the finite computational step size of \textsc{Mathematica}. Nevertheless this effect, as measured by an Unruh-deWitt detector, seems to be the first of its kind. Indeed it is comparable to the result obtained for the BTZ black hole; however in that case, the transition probability exhibited \textit{continuous} resonances at special values of the black hole mass ratio. 

\begin{figure}[h]
    \centering
    \includegraphics[width=0.9\linewidth]{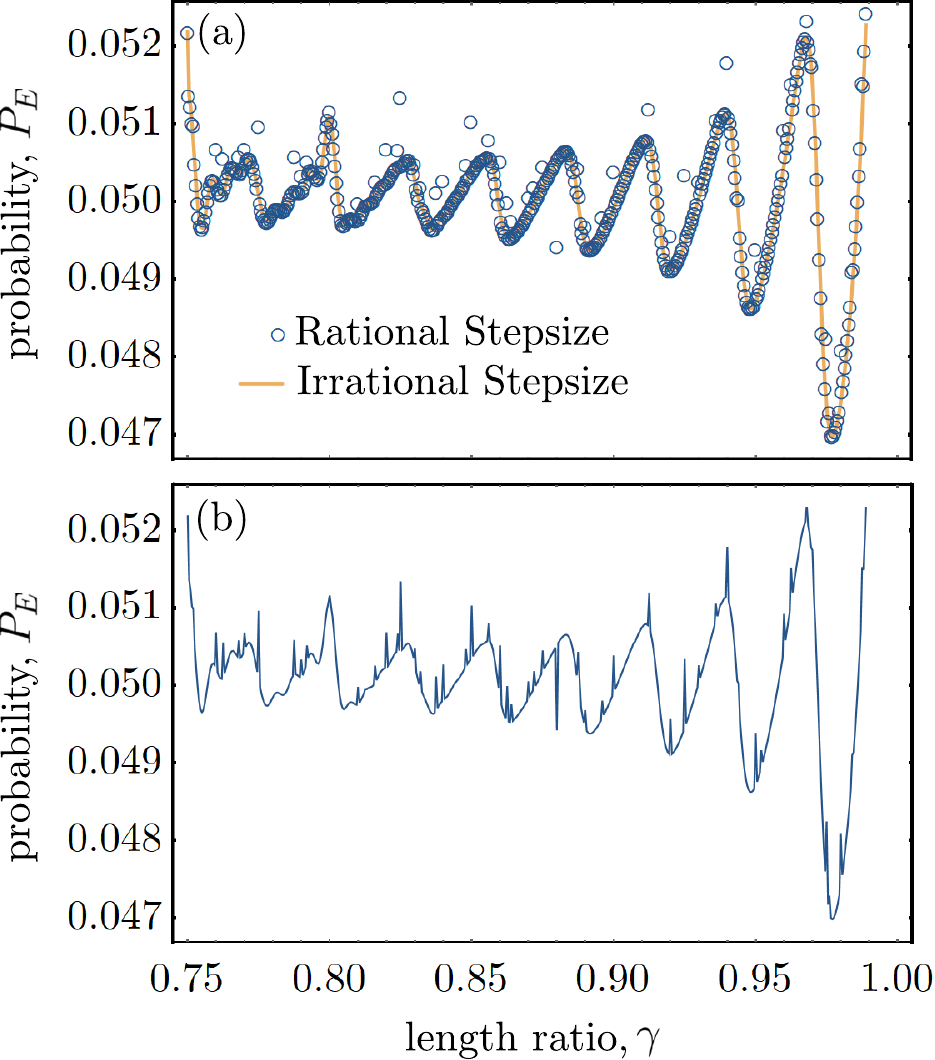}
    \caption{(a) Illustration of the discontinuous behaviour of the transition probability as a function of $\gamma$. The continuous yellow line was plotted with an irrational stepsize, and hence the resonant peaks at rational values of $\gamma$ do not appear. (b) The same dataset as shown with the open circles in (a), but with the data points connected. We have used the settings $\Omega \sigma = 1/100$ and $l_A = 2$. }
    \label{fig:2}
\end{figure}

To illustrate this resonant effect further, we have plotted the transition probability of the detector as a function of $\gamma$ using rational and irrational step sizes in Fig.\ \ref{fig:2}. When the step size used to plot $P_E$ is irrational, the transition probability appears to be smooth and continuous. This strongly contrasts the discontinuous nature of $P_E$ when utilizing rational step sizes in \textsc{Mathematica}, and thus confirms the source of the resonant effect shown in Fig.\ \ref{fig:PE}.

\begin{figure}[h]
    \centering
    \includegraphics[width=0.9\linewidth]{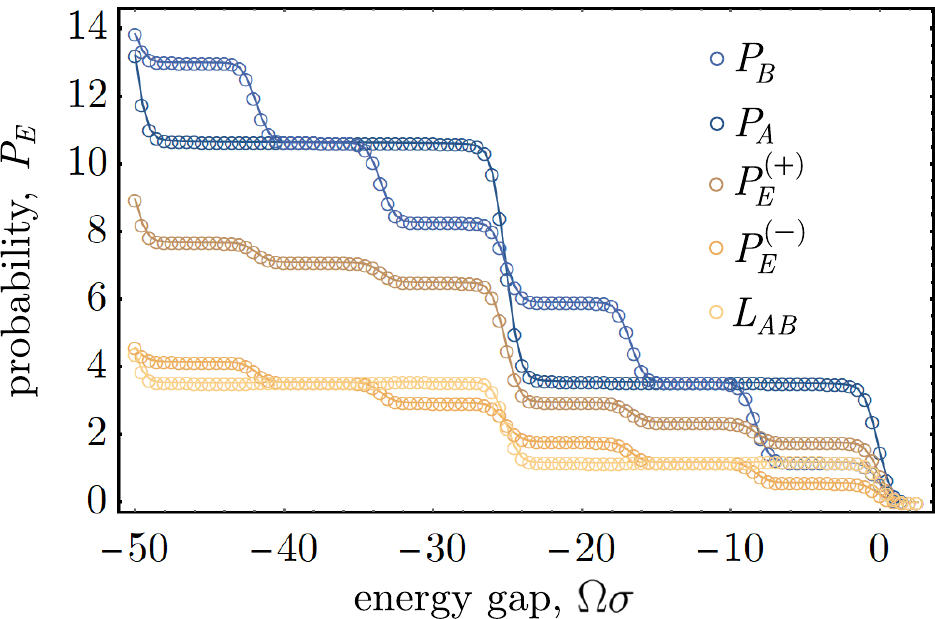}
    \caption{Transition probability of the detector as a function of the energy gap. The different colours correspond to different contributions to the transition probability (dark blue, light blue, light yellow) and the total transition probability upon conditioning in the $| \pm \rangle$ state (brown, orange). We have chosen $l_A/\sigma = 0.25$, $l_B/\sigma = 0.75$. }
    \label{fig:3}
\end{figure}

In Fig.\ \ref{fig:3}, we have plotted the transition probability as a function of the energy gap of the detector, for a superposition of two lengths $l_A$, $l_B$. The individual contributions to the transition probability are shown in different colours, giving the total result displayed with the dashed lines (the different colours representing two different measurement bases for the control). As explained in \cite{martinmartinezPhysRevD.93.044001}, the oscillations in the individual contributions with $\Omega$ is akin to the appearance of modified quasinormal modes in spacetimes with closed topology. The cross-term $L_{AB}$ also exhibits similar behaviour, an expression of the fact that it encodes the quantum interference between the two topologically identified spacetimes. 

% \begin{figure}[htp]
%     \begin{subfigure}{0.5\textwidth}
%         \includegraphics[width=0.5\textwidth]{coefficient LA 1to6.PNG}
%         \label{fig:my_label}
%         \caption{$\gamma=\frac{l_A}{l_B}\geq 0$ on x axis and total transition probability on the y axis.}
%     \end{subfigure}\vspace{2mm}
%     \hfill
%     \begin{subfigure}{0.5\textwidth}
%         \includegraphics[width=0.5\textwidth]{picof fixed LA small coefficent.PNG}
%         \label{fig:my_label}
%         \caption{$\gamma=\frac{l_A}{l_B}\leq 0$ on x axis and total transition probability on the y axis.}
%     \end{subfigure}
%     \caption{Here, at the above plots, we show how transition probability varies with different $\frac{l_A}{l_B}$ ratios where $l_A$ changes and $l_B$ kept constant. The variations correlate with the coefficient function since it is a factor of the Minkowski and second term, a closed space with identifications (See appendix for the correlation values that correspond to the figure 1 and 2 $l_A$ and $l_B$ values.) }
%     \end{figure}

% \includegraphics[scale=0.5]{LAtoLB_small_LA_plot.pdf}

\textit{Analogy with Cavity QFT}\textemdash 
In the previous section, we looked at a scalar field quantised on a %the quantum-gravitational  effects induced by a
(3+1)-dimensional spacetime with a periodic boundary condition (along one spatial dimension), where the length of this dimension depends on a quantum degree of freedom. We considered effects that arise when this degree of freedom is in a superposition, meaning the spacetime is itself in superposition of different sizes. We found analogous effects to those arising in a BTZ  black hole spacetime where the mass of the black hole is in a corresponding superposition.   %which we consider to be in a balanced superpositionm  topologica length is superposition coupled to a two-level particle detector via a massless scalar field. 
This strengthens the case for physical relevance of the associated effects and it also makes it worthwhile to look for a possible experimental realisation of an analog system which would help illuminate the role and physical consequences of our physical assumptions. These include the choice of an initial state and the role of potential nontrivial free dynamics of the  degree of freedom associated with the spacetime.

Below we outline one idea for such an analog and discuss what are the outstanding conceptual challenges. Note that quantum fields in the spacetime considered in the present work share similarities to a quantum field with periodic boundary condition -- imposed in a quantum superposition of characteristic lengths.\footnote{Henceforth, we refer to the periodically identified field in (1+1) as a ``cavity'', although strictly speaking it more accurately mimics a field quantized on a ring resonator-type potential \cite{bowen2015quantum}.} Thus, we focus here on such a cavity system.

For concreteness, let us consider three mode decompositions $\hat{\phi}_A(t,x)$, $\hat{\phi}_B(t,x)$, $\hat{\phi}_C(t,x)$ of a (1+1)-dimensional Klein-Gordon field:
\begin{align}
    \hat{\phi}_A(t,x) &= \sum_{n\neq0} \Big( f_n(t,x) \hat{a}_k + \hat{f}_n^\star(t,x) \hat{a}_k^\dd \Big) ,
    \\
    \hat{\phi}_B(t,x) &= \sum_{m\neq0} \Big( g_m(t,x) \hat{b}_\omega + g_m^\star(t,x) \hat{b}_\omega^\dd \Big) ,
    \\
    \hat{\phi}_C(t,x) &= \sum_{l \neq0} \Big( h_l(t,x) \hat{c}_\Omega + h_l^\star(t,x) \hat{c}_\Omega^\dd \Big) ,
\end{align}
where the mode functions are defined as
\begin{align}
    f_n(t,x) &= \frac{1}{\sqrt{4\pi|n|}} e^{-i|k_n|t + i k_nx} , 
    \\
    g_m(t,x) &= \frac{1}{\sqrt{4\pi|m|}} e^{-i|\omega_m|t + i\omega_mx} , 
    \\
    h_l(t,x) &= \frac{1}{\sqrt{4\pi |l|}} e^{-i|\Omega_l|t + i \Omega_lx} , 
\end{align}
and the field momenta take on discrete values, $k_n = (2\pi n)/l_A$, $\omega_m = (2\pi m)/l_B$ and $\Omega_l = (2\pi l )/l_C$. Moreover, $n,m,l$ are nonzero integers (we neglect the infamous ``zero'' mode, which requires a separate treatment \cite{tjoaPhysRevD.101.125020,tjoaPhysRevD.99.065005,martinmartinezPhysRevD.93.044001}).

In our Minkowski spacetime superposition, the correlation functions are not evaluated with respect to the ``local vacua'' associated with the identifications $l_A$, $l_B$, but rather with the ``global'' Minkowski vacuum state $|0\rangle \equiv | 0_M\rangle $. This is the reason we consider three sets of modes above, where the set $\{\hat{c}_\Omega, \hat{c}_\Omega^\dd\}$ serves as the analog of the global modes, and we denote the associated global vacuum $|0_C\rangle$ (which is a state annihilated by $\hat{c}_\Omega$). Similar decompositions have been considered in the context of entanglement production between the local modes inside a larger ``global'' cavity \cite{brownPhysRevD.91.016005,vazquez2014local}. In order to evaluate the required Wightman functions, we next express the operators $(\hat{a}_k, \hat{b}_\omega)$ associated with the local modes in terms of those associated with the global modes $\hat{c}_\Omega$. This is done using Bogoliubov transformations between the modes:
\begin{align}
    \hat{a}_k &= \sum_{l\neq0} \Big( \alpha_{k\Omega} \hat{c}_\Omega + \beta_{k\Omega} \hat{c}_\Omega^\dd \Big) 
    \\
    \hat{b}_\omega &= \sum_{l\neq0} \Big( \bar{\alpha}_{k\Omega} \hat{c}_\Omega + \bar{\beta}_{k\Omega} \hat{c}_\Omega \Big) 
\end{align}
where $\alpha_{k\Omega} = \langle f_n, h_l \rangle$, $  \beta_{k\Omega} = \langle f_n^\star, h_l \rangle$, $   \bar{\alpha}_{\omega\Omega} = \langle g_m , h_l \rangle$
$\bar{\beta}_{\omega\Omega} = \langle g_m^\star, h_l \rangle$, and where the Klein-Gordon inner product is defined in the usual way \cite{birrell1984quantum} 
\begin{align}
    \langle \phi_1, \phi_2 \rangle &= i \int_V \D x \: ( \phi_1^\star \p_t \phi_2 - \phi_2 \p_t \phi_1^\star ).
\end{align}
% where the Bogoliubov coefficients are given by 
% \begin{align}
%     \alpha_{k\Omega} &= i \int_V \D x \: \Big( f_n^\star \p_t h_l - h_l \p_t f_n^\star \Big) 
%     \\
%     \beta_{k\Omega} &= i \int_V \D x \: \Big( f_n^\star \p_t h_l^\star - h_l^\star \p_t f_n^\star \Big) 
%     \\
%     \bar{\alpha}_{\omega\Omega} &= i \int_V \D x \: \Big( g_m^\star \p_t h_l - h_l \p_t g_m^\star \Big) 
%     \\
%     \bar{\beta}_{\omega\Omega} &= i \int_V \D x \: \Big( g_m^\star \p_t h_l^\star - h_l^\star \p_t g_m^\star \Big)
% \end{align}

%For a detector residing in a superposition of topologically identified Minkowski spacetimes, its transition probability was found to be the sum of four terms: two contributions from the individual amplitudes of the spacetime superposition, and two cross-correlation terms between them. This will also be the case here. 
The key structures relevant to our analysis are Wightman functions evaluated for the products of fields with the same characteristic length $l_{A(B)}$ (``local terms'') and with different lengths (``cross terms''). Let us examine the functional form of these terms, (where for brevity and without loss of generality, we display ``local'' terms for the cavity with length $l_A$). 

The local term reads
\begin{align}
    W(\mtx_A, \mtx_A') &= \sum_{i} \sum_{n,m,l} W_i (\mtx_A, \mtx_A') 
\end{align}
%with the individual terms taking the form, 
% \begin{align}
%     W_1(\mtx_A, \mtx_A') &= \sum_{n,m,l} \frac{e^{-i|k_n|t-i|k_m|t'}}{4\pi \sqrt{|n||m|}}  \alpha_{k\Omega} \beta_{k'\Omega} 
%     \\
%     W_2 (\mtx_A, \mtx_A') &= \sum_{n,m,l} \frac{e^{-i|k_n|t + i|k_m|t'}}{4\pi \sqrt{|n||m|}}  \alpha_{k\Omega} \alpha_{k'\Omega}^\star
%     \\
%     W_3(\mtx_A, \mtx_A') &= \sum_{n,m,l} \frac{e^{i|k_n|t - i|k_m|t'}}{4\pi \sqrt{|n||m|}}  \beta_{k\Omega}^\star \beta_{k'\Omega} 
%     \\
%     W_4 (\mtx_A, \mtx_A') &=\sum_{n,m,l} \frac{e^{i|k_n|t + i|k_m|t'}}{4\pi \sqrt{|n||m|}}  \beta_{k\Omega}^\star \alpha_{k'\Omega}^\star 
% \end{align}
% which gives, 
where 
\begin{align}
    W_1(\mtx_A, \mtx_A') &= - \Omega_{k_n}^{1} \Omega^{2}_{k_m} \Delta_{nl;k}^{1} \Delta_{ml;k}^{2} ,
    \vt 
    \\
    W_2( \mtx_A, \mtx_A') &= \Omega^{1}_{k_n} \Omega^{1}_{k_m} \Delta_{nl;k}^{1} \Delta_{ml;k}^{1\star}
    \vt ,
    \\
    W_3( \mtx_A, \mtx_A' ) &= \Omega^{2}_{k_n} \Omega^{2}_{k_m} \Delta_{nl;k}^{2\star} \Delta_{ml;k}^{2}
    \vt , 
    \\
    W_4 (\mtx_A, \mtx_A') &= - \Omega^{2}_{k_n} \Omega^{1}_{k_m} \Delta_{nl;k}^{2\star} \Delta_{ml;k}^{1\star} 
    \vt ,
\end{align}
and 
\begin{align}
    \Omega^{1}_{k_j} &= \Big( |\Omega_l| + |k_j| \Big) \frac{e^{-i|\Omega_l|s/2}}{(4\pi)^{3/2}\sqrt{|n||m||l|}}
    \vphantom{ \begin{cases}
    - i l_A & k_j = \Omega_l 
    \\
    \frac{1 - e^{i(\Omega_l-k_j)l_A}}{\Omega_l-k_j} & k_j \neq \Omega_l 
    \end{cases}}
    \\
    \Omega^{2}_{k_j} &= \Big( |\Omega_l| - |k_j| \Big) \frac{e^{-i|\Omega_l|s/2}}{(4\pi)^{3/2}\sqrt{|n||m||l|}}
    \vphantom{ \begin{cases}
    - i l_A & k_j = \Omega_l 
    \\
    \frac{1 - e^{i(\Omega_l-k_j)l_A}}{\Omega_l-k_j} & k_j \neq \Omega_l 
    \end{cases}}
    \\
    \label{eq48}
    \Delta_{jl;k}^{1} &= \begin{cases}
    - i l_A & k_j = \Omega_l 
    \\
    \frac{1 - e^{i(\Omega_l-k_j)l_A}}{\Omega_l-k_j} & k_j \neq \Omega_l 
    \end{cases}
    \\
    \Delta_{jl;k}^{2} &= \begin{cases}
    - il_A & k_j = - \Omega_l 
    \\
    \label{eq49}
    \frac{1 - e^{-i(\Omega_l + k_j)l_A}}{\Omega_l + k_j} & k_j \neq - \Omega_l 
    \end{cases}
\end{align}
Our nomenclature in Eq.\ (\ref{eq48}) and (\ref{eq49}) is chosen to highlight the Kronecker-$\delta$-like property of these functions, which ``pull out'' a term proportional to $l_A$ at certain resonant values of the cavity lengths. This property shows the first conceptual challenge: as we consider a large but finite global cavity, the terms in the Wightman function sums where $k_j = \Omega_l$ will lead to resonances independently of the ratio between the superposed lengths (not directly given by $l_{A}/l_{B}$) -- a direct consequence of the present toy model. %A possible way out is to assume that length $l_C$ is such that neither $l_A/l_C$ nor $l_B/l_C$ are rational.

% It can be shown numerically that the expression for $W(\mtx_A, \mtx_A')$ reduces to the `standard' single cavity limit as $l_A \to l_C$:
% \begin{align}
%     \lim_{l_A \to l_C} W(\mtx_A, \mtx_A') &= \sum_{n\neq0} \frac{e^{-i|k_n|s}}{4\pi |n|} 
% \end{align} APPENDIX!
The cross-term follows analogously, 
\begin{align}
    W(\mtx_A, \mtx_B') &= \sum_i \sum_{n,m,l} W_i(\mtx_A, \mtx_B') 
\end{align}
where
\begin{align}
    W_1(\mtx_A, \mtx_B') &= - \Omega^{1}_{k_n} \Omega^{2}_{\omega_m} \Delta_{nl;k}^1 \Delta_{ml;\omega}^2
    \vt , 
    \\
    W_2( \mtx_A, \mtx_B') &=  \Omega_{k_n}^{1} \Omega_{\omega_m}^{1} \Delta_{nl;k}^{1} \Delta_{ml;\omega}^{1\star} 
    \vt , 
    \\
    W_3( \mtx_A, \mtx_B' ) &= \Omega_{k_n}^{2} \Omega_{\omega_m}^{2} \Delta_{nl;k}^{2\star} \Delta_{ml;\omega}^{2}
    \vt ,
    \\
    W_4 (\mtx_A, \mtx_B') &= - \Omega_{k_n}^{2} \Omega_{\omega_m}^{1}
    \Delta_{nl;k}^{2\star} \Delta_{ml;\omega}^{1\star} 
    \vt ,
\end{align}
and
\begin{align}\label{67}
    \Delta_{jl;\nu}^{1} &= \begin{cases}
    - i l_i & \nu_j = \Omega_l 
    \\
    \frac{1 - e^{i(\Omega_l-\nu_j)l_i}}{\Omega_l-\nu_j} & \nu_j \neq \Omega_l 
    \end{cases}
    \\
    \label{68}
    \Delta_{jl;\nu}^{2} &= \begin{cases}
    - il_i & \nu_j = - \Omega_l 
    \\
    \frac{1 - e^{-i(\Omega_l + \nu_j)l_i}}{\Omega_l + \nu_j} & \nu_j \neq - \Omega_l 
    \end{cases}
\end{align}
and in the above $l_i = l_A$ if $\nu = k$ and $l_i = l_B$ if $\nu = \omega$. 

Equations (\ref{67}) and (\ref{68}) are the main results of this section. As with the local term, the $\Delta$-functions have properties similar to the Kronecker-$\delta$, selecting unique terms when the cavity length $l_A$ resonates with the global cavity $l_C$, and likewise when the cavity length $l_B$ resonates with $l_C$. %We thus expect that a UdW-type detector situated inside such a superposed cavity will experience discrete resonances in its transition probability at rational ratios of the superposed cavity lengths. 

The above analysis highlights that the analogy between the cavity model and the Minkowski cylinder spacetime is not perfect. This is due to the following reasons: Firstly, we have utilized a discrete mode decomposition of the field, which, upon neglecting the ``zero-mode'', yields a slightly different limit to the spacetime example. Secondly, the spacetime example considers the ``global'' vacuum state $|0 \rangle$ as being defined with respect to an infinite (not periodically identified) Minkowski spacetime, that is, in the limit of $l \to \infty$. Such a limit is not well-defined in the cavity example, for this would lead to a continuum of $\textbf{k}$ vectors describing the global modes and a discrete quantization for the local modes. A result of the discreteness is that not only do the individual amplitudes of the cavity resonate with each other (i.e.\ the presence of $\Delta$-functions in $\Omega$, $\nu$ in Eq.\ (\ref{67}) and (\ref{68}) which characterize the respective lengths of the superposed cavity), but the individual cavities resonate with the large, yet finite ``global mode'' cavity. %This feature could complicate the interpretation of unambiguously identifying a resonance effect arising from interference between the superposed cavity lengths. 
A possible way out could be to choose the global cavity length $l_C$ such that neither $l_A/l_C$ nor $l_B/l_C$ are rational and examine the resulting resonances and their dependence on $l_A/l_B$.

Strong motivation for further study of this system is that it could be possible to simulate it using optomechanical technologies. For example, experiments have realized superfluid Helium condensates on toroidal cavities. The fundamental mode providing the ``analog metric'' for phonons in the condensate therein can interact with incident photons, thus suggesting an opportunity to study ``light-controlled'' effective quantum-superposed backgrounds \cite{bowen2015quantum}.

\textit{Conclusions}\textemdash In this paper, we have studied %via operational methods, 
the response of a two-level quantum system (a UDW detector) to a massless scalar field on a background Minkowski spacetime in a superposition of topologically nontrivial identifications. The topology-superposition of the background spacetime elicits resonances in the particle's transition probability when the ratio of periodic lengths of the spacetimes in superposition is a rational value. This effect corroborates a related result obtained recently for the (2+1)-dimensional black hole in a superposition of masses. Our results further highlight 
the importance  of the choice of the vacuum state (of the field) in the present and similar models of spacetime superpositions. In the present work we have made the simple choice of the global Minkowski vacuum. 

We have also investigated the potential for a cavity realisation of such a spacetime superposition -- modelling the periodically identified Minkowski quotient space as a quantum field with periodic boundary condition in a cavity. While further refinements of this model are needed, the cavity model represents an interesting direction for further study.

% \rule{8.5cm}{0.02cm}
\newpage 

%[x] Eduardo Martin-Martinez, Alexander R. H. Smith, and Daniel R. Terno Space-time structure and vacuum entanglement, Physical Review D 93, 044001 (2016) DOI: 10.1103/PhysRevD.93.044001 \\
%[??] A. Kempf, Replacing the notion of spacetime distance by the notion of correlation, Frontiers 11 in Physics 9, 10.3389/fphy.2021.655857 (2021).
\begin{widetext}
\section{Appendix}
\subsection{Transition probability with double-sum Wightman functions}
In this section, we derive the analytical expressions for the ``local contributions'' to the transition probability. Recall that the Wightman function take the form 
\begin{align}
    W_{J_0} (\mtx, \mtx' ) &= \frac{1}{\mathcal{N}} \sum_{n,m} \bigg[ \frac{\mathrm{sgn}(\tau - \tau') \delta( ( \tau - \tau')^2 - l^2(n-m)^2 ) }{4\pi i } - \frac{1}{4\pi^2( ( \tau - \tau'^2 - l^2( n-m)^2)} \bigg] \non \\
    &= \frac{1}{\mathcal{N}} \sum_{n,m} \bigg[ \frac{\mathrm{sgn}(s) \delta(s^2 - l^2(n-m)^2)}{4\pi i }  - \frac{1}{4\pi^2 ( s^2 - l^2(n-m)^2) } \bigg] \non \\
    &= \frac{1}{\mathcal{N}} \sum_{n=m} \bigg[ \frac{\mathrm{sgn}(s)\delta(s^2)}{4\pi i } - \frac{1}{4\pi^2s^2} \bigg] + \frac{1}{\mathcal{N}} \sum_{n\neq m} \bigg[ \frac{\mathrm{sgn}(s) \delta(s^2 - l^2(n-m)^2)}{4\pi i } - \frac{1}{4\pi^2(s^2-l^2(n-m)^2)} \bigg] \non \\
    &= W_M(s) + \frac{1}{\mathcal{N}} \sum_{n\neq m } \bigg[ \frac{\mathrm{sgn}(s) \delta(s^2 - l^2(n-m)^2)}{4\pi i} - \frac{1}{4\pi^2(s^2 - l^2(n-m)^2)} \bigg] 
\end{align}
where we have defined $s = \tau - \tau'$ as the proper time difference. The transition probability is given by 
\begin{align}
    P_D &= \infint\D \tau \infint\D \tau' e^{- \frac{\tau^2}{2\sigma^2}} e^{- \frac{\tau'^2}{2\sigma^2}} e^{-i\Omega ( \tau - \tau' ) } W_{J_0}(\mtx, \mtx' ) 
    \\
    &= \infint\D u \infint\D s \: e^{- \frac{u^2}{2\sigma^2}} e^{- \frac{(u-s)^2}{2\sigma^2}} e^{-i\Omega s } W_{J_0}(\mtx ,\mtx' ) \\
    &= \sqrt{\pi}\sigma \infint\D s \: e^{- \frac{s^2}{4\sigma^2}} e^{-i\Omega s} W_{J_0}(s) ,
\end{align}
having performed the $\D u$ integral in the last line. Thus, 
\begin{align}
    P_D &= \sqrt{\pi}\sigma \infint\D s \: e^{- \frac{s^2}{4\sigma^2}} e^{-i\Omega s} \bigg[  W_M(s) + \frac{1}{\mathcal{N}}\sum_{n\neq m } \bigg[ \frac{\mathrm{sgn}(s) \delta(s^2 - l^2(n-m)^2)}{4\pi i } - \frac{1}{4\pi^2(s^2 - l^2(n-m)^2)} \bigg] \bigg]  \non  \\
    &= P_M + \frac{\sqrt{\pi }\sigma }{\mathcal{N}} \sum_{n\neq m } \infint\D s \: e^{- \frac{s^2}{4\sigma^2}} e^{-i\Omega s}  \bigg[ \underbrace{\frac{\mathrm{sgn}(s) \delta(s^2 - l^2(n-m)^2)}{4\pi i }}_{I_1} - \underbrace{\frac{1}{4\pi^2(s^2 - l^2(n-m)^2)}}_{I_2} \bigg] 
\end{align}
Let us examine the two terms in the image sum. We have
\begin{align}
    I_1 &= \frac{1}{4\pi i} \sum_{n\neq m} \infint\D s \: e^{- \frac{s^2}{4\sigma^2}} e^{-i\Omega s } \mathrm{sgn}(s) \delta(s^2 - l^2(n-m)^2) \\
    &= \frac{1}{4\pi i } \sum_{n\neq m}\infint\D s \: e^{- \frac{s^2}{4\sigma^2}} e^{-i\Omega s } \mathrm{sgn}(s)
    \frac{1}{2|l(n-m) |} \Big[  \delta(s + l(n-m) ) + \delta(s - l(n-m) ) \Big] \non 
    \\
    &=  \frac{1}{4\pi i } \sum_{n\neq m} \frac{e^{- \frac{l^2(n-m)^2}{4\sigma^2}}}{2|l(n-m)|} \bigg[ e^{i \Omega l(n-m)}\mathrm{sgn}(-l(n-m)) + \mathrm{sgn}(l(n-m)) e^{-i\Omega l(n-m)} \bigg] \non
\intertext{It is convenient here to split up the sum into contributions where $n>m$ and $n<m$, yielding:}
    I_1 &= \frac{1}{4\pi i } \sum_{n>m} \frac{e^{-\frac{l^2(n-m)^2}{4\sigma^2}}}{2l|n-m|} \bigg[ -e^{i\Omega l (n-m)} \mathrm{sgn}(l(n-m)) + \mathrm{sgn}(l(n-m)) e^{-i\Omega l(n-m)} \bigg] \non \\
    & + \frac{1}{4\pi i } \sum_{m>n} \frac{e^{-\frac{l^2(n-m)^2}{4\sigma^2}}}{2l|n-m|} \bigg[ e^{i\Omega l (n-m) } \mathrm{sgn}(l(m-n) ) - e^{-i\Omega l (n-m) } \mathrm{sgn}(l(m-n ) ) \bigg] \non  \\
    &= \frac{1}{4\pi l} \sum_{n>m} \frac{e^{-\frac{l^2(n-m)^2}{4\sigma^2}}}{n-m} \sin(\Omega l (m-n ) ) - \frac{1}{4\pi l} \sum_{m>n} \frac{e^{-\frac{l^2(n-m)^2}{4\sigma^2}}}{m-n} \sin(\Omega l(m-n ) ) \non \\
    &= \frac{1}{2\pi l} \sum_{n>m} \frac{e^{-\frac{l^2(n-m)^2}{4\sigma^2}}}{n-m}\sin(\Omega l (m-n)) 
    = - \frac{1}{2\pi l } \sum_{n>m} \frac{e^{- \frac{l^2(n-m)^2}{4\sigma^2}}}{n-m} \sin(\Omega l(n-m)) .
\end{align}
For the second term, we have 
\begin{align}
    I_2 &= - \frac{1}{4\pi^2} \sum_{n\neq m} \infint\D s \: \frac{e^{-\frac{s^2}{4\sigma^2}}e^{-i\Omega s}}{s^2-l^2(n-m)^2} \\
    &= - \frac{1}{4\pi^2} \sum_{n\neq m} \infint\D s \: \infint\D s' \delta(s - s') \frac{e^{- \frac{s'^2}{4\sigma^2}} e^{-i\Omega s'}}{s^2-l^2(n-m)^2} \\
    &= - \frac{1}{4\pi^2} \sum_{n\neq m } \infint\D s \infint\D s' \left( \frac{1}{2\pi} \infint\D z \: e^{iz(s' - s) } \right) \frac{e^{- \frac{s'^2}{4\sigma^2}}e^{-i\Omega s'} }{s^2-l^2(n-m)^2} \\
    &= - \frac{1}{8\pi^3} \sum_{n\neq m} \infint\D z \left( \infint\D s' e^{-(\Omega - z)s'} e^{- \frac{s'^2}{4\sigma^2}} \right) \left( \infint\D s \: \frac{e^{-isz} }{s^2-l^2(n-m)^2} \right)  \\
    &= - \frac{1}{8\pi^3} \sum_{n\neq m } \infint\D z\left( 2 \sqrt{\pi}\sigma e^{-(\Omega - z)^2\sigma^2} \right) \left( - \pi \mathrm{sgn}(z) \frac{\sin ( l (n-m) z)}{l (n-m) } \right)  
\intertext{Performing the integration over $z$ leaves the compact analytic expression,}
    I_2 &=  \sum_{n\neq m} \frac{e^{-\frac{l^2(n-m)^2}{4\sigma^2}}}{4\pi l (n-m)}  \mathrm{Im} \bigg[ e^{i l(n-m) \Omega} \mathrm{erf} \left( \frac{il(n-m)}{2\sigma} + \sigma \Omega \right) \bigg]. 
\end{align}
The total expression for the transition probability is thus 
\begin{align}
    P_D &= P_M + \frac{\sigma}{2 \sqrt{\pi} l\sum_n \eta^{2n}}  \sum_{n\neq m} \frac{e^{-\frac{l^2(n-m)^2}{4\sigma^2}}}{2(n-m)} \mathrm{Im} \left[ e^{il(n-m)\Omega} \mathrm{erf} \left( \frac{il(n-m)}{2\sigma} + \sigma\Omega \right) \right] 
    \non 
    \\
    \label{eq69}
    & - \frac{\sigma}{2 \sqrt{\pi} l\sum_n \eta^{2n}} \sum_{n>m} \frac{e^{-\frac{l^2(n-m)^2}{4\sigma^2}}}{n-m} \sin ( \Omega l (n-m) ) ,
\end{align}
as stated in the main text. It can be straightforwardly verified that this gives an identical result to that obtained using only a single sum expression for the Wightman function, 
\begin{align}\label{eq70}
    P_D &= P_M + \frac{\sigma}{2\sqrt{\pi}} \sum_{n=1}^\infty \frac{e^{-\frac{l^2(n-m)^2}{4\sigma^2}}}{l n} \left( \mathrm{Im} \left[ e^{il n\Omega} \mathrm{erf}\left( \frac{iln}{2\sigma}  + \sigma\Omega \right) \right] - \sin ( \Omega l n ) \right) .
\end{align}

\subsection{Cross-correlation term with double-sum Wightman functions}
The cross-term can be similarly derived. Inserting the cross-correlation Wightman function into the integral expressions for $L_{AB}$ yields
\begin{align}
    L_{AB} &= \frac{\sqrt{\pi}\sigma}{\mathcal{N}} \sum_{n,m} \infint\D s \: e^{- \frac{s^2}{4\sigma^2}} e^{-i\Omega s} \bigg[ \frac{\mathrm{sgn}(s) \delta( s^2 - (l_A n - l_B m)^2)}{4\pi i} - \frac{1}{4\pi^2 ( s^2 - (l_An - l_Bm)^2} \bigg] \non \\
    &= \frac{\sqrt{\pi}\sigma}{\mathcal{N}} \sum_{l_An = l_B m} \infint\D s \: e^{- \frac{s^2}{4\sigma^2}} e^{-i\Omega s} \bigg[ \frac{\mathrm{sgn}(s) \delta(s^2 ) }{4\pi i } - \frac{1}{4\pi^2s^2} \bigg] \non \\
    & + \frac{\sqrt{\pi}\sigma}{\mathcal{N}} \sum_{l_An \neq l_Bm } \infint\D s \: e^{- \frac{s^2}{4\sigma^2}} e^{-i\Omega s} \bigg[ \frac{\mathrm{sgn}(s) \delta( s^2 - (l_An - l_B m )^2)}{4\pi i } - \frac{1}{4\pi^2(s^2 - (l_A n - l_B m )^2)} \bigg] \non  \\
    &= \frac{1}{\mathcal{N}} \sum_{l_An=l_Bm} P_M + \frac{\sqrt{\pi}\sigma}{\mathcal{N}} \sum_{l_An \neq l_B m } \infint\D s \: e^{- \frac{s^2}{4\sigma^2}} e^{-i\Omega s} \bigg[ \frac{\mathrm{sgn}(s) \delta(s^2 - (l_An -l_Bm)^2)}{4\pi i } \non \\
    & - \frac{1}{4\pi^2(s^2 - (l_An-l_Bm)^2)} \bigg] 
\end{align}
Let's look at the image sum integrals. We have,
\begin{align}
    I_1 &= \sum_{l_An\neq l_Bm} \infint\D s \: e^{- \frac{s^2}{4\sigma^2}}e^{-i\Omega s} \frac{\mathrm{sgn}(s) \delta( s^2 - (l_An-l_Bm)^2)}{4\pi i } \\
    &= \sum_{l_An\neq l_Bm} \infint\D s \: e^{- \frac{s^2}{4\sigma^2}} e^{-i\Omega s} \frac{\mathrm{sgn}(s) \delta(s^2 - (l_A n - l_Bm)^2)}{4\pi i } \\
    &= \frac{1}{4\pi i }\sum_{l_An\neq l_Bm} \infint\D s \: e^{- \frac{s^2}{4\sigma^2}} e^{-i\Omega s } \mathrm{sgn}(s) \frac{1}{2|l_An - l_Bm|} \Big[ \delta(s + l_An - l_Bm ) + \delta( s - l_An + l_Bm) \Big] 
    \\
    &= \frac{1}{4\pi i }\sum_{l_An \neq l_B m } \bigg[ \frac{e^{- \frac{(l_A n - l_Bm)^2}{4\sigma^2}} e^{i\Omega( l_A n - l_Bm)} \mathrm{sgn}(-l_An +l_Bm)}{2| l_A n - l_B m | } + \frac{e^{- \frac{(l_A n - l_B m)^2}{4\sigma^2}} e^{-i \Omega( l_A n - l_B m)} \mathrm{sgn}( l_A n - l_B m ) }{2| l_A n - l_B m |} \bigg] 
\end{align}
We can split up the summation 
\begin{align} 
    I_1 &= \frac{1}{4\pi i } \sum_{l_An > l_B m} e^{-\frac{(l_A n - l_Bm)^2}{4\sigma^2}} \bigg[ \frac{e^{i\Omega( l_An - l_Bm)} \mathrm{sgn}(-l_An+l_Bm)}{2|l_An-l_Bm|} + \frac{e^{-i\Omega(l_An-l_Bm)} \mathrm{sgn}(l_An-l_Bm)}{2|l_An-l_Bm|} \bigg] \non \\
    & + \frac{1}{4\pi i } \sum_{l_Bm > l_An} e^{-\frac{(l_A n - l_Bm)^2}{4\sigma^2}} \bigg[ \frac{e^{i\Omega(l_A n - l_Bm)}\mathrm{sgn}(-l_A n + l_Bm)}{2|l_An-l_Bm|} + \frac{e^{-i\Omega( l_An - l_Bm)} \mathrm{sgn}(l_A n - l_Bm)}{2|l_An - l_Bm|} \bigg] \\
    %%%%
    &= \frac{1}{4\pi } \bigg[ \sum_{l_An > l_B m} \frac{e^{- \frac{(l_An-l_Bm)^2}{4\sigma^2}}}{l_An-l_Bm}  \sin(\Omega(l_Bm-l_An)) - \sum_{l_Bm > l_An} \frac{e^{- \frac{(l_An - l_Bm}{4\sigma^2}}}{l_Bm-l_An}  \sin(\Omega(l_Bm-l_An))  \bigg] \non \\
    &= \frac{1}{2\pi } \sum_{l_A n > l_Bm} \frac{e^{- \frac{(l_A n - l_B m)^2}{4\sigma^2}}}{l_A n - l_B m } \sin ( \Omega ( l_B m - l_A n ) ) = - \frac{1}{2\pi} \sum_{l_An>l_Bm} \frac{e^{- \frac{(l_An-l_Bm)^2}{4\sigma^2}}}{l_An-l_Bm} \sin(\Omega(l_A n - l_B m ) ) ,
\end{align}
in close analogy to the local contributions to the transition probability. Next,
\begin{align}
    I_2 &=  \frac{1}{4\pi^2} \sum_{l_A n \neq l_B m } \infint\D s \: \frac{e^{- \frac{s^2}{4\sigma^2}} e^{- i\Omega s}}{s^2 - (l_A n - l_B m )^2} \\
    &= \frac{\sqrt{\pi}\sigma}{\sum_n \eta^{2n}} \frac{1}{4\pi^2} \sum_{l_An\neq l_B m } \infint\D s \infint\D s' \delta(s-s')  \frac{e^{-\frac{s'^2}{4\sigma^2}} e^{-i\Omega s'}}{s^2-(l_An-l_Bm)^2} \\
    &= \frac{1}{4\pi^2} \sum_{l_An\neq l_Bm} \infint\D s \infint\D s' \left( \frac{1}{2\pi} \infint\D z \: e^{iz(s'-s)} \right) \frac{e^{-\frac{s'^2}{4\sigma^2}} e^{-i\Omega s'}}{s^2 - (l_A n - l_B m )^2} \non \\
    &= \frac{1}{8\pi^3} \sum_{l_A n \neq l_B m } \infint\D z \left( \infint\D s' e^{-(\Omega - z)^2 s'} e^{- \frac{s'^2}{4\sigma^2}} \right) \left( \infint\D s \: \frac{e^{-isz}}{s^2 - (l_An - l_Bm)^2} \right) \non \\
    &= \frac{1}{8\pi^3} \sum_{l_A n \neq l_B m } \infint\D z \: \left( 2 \sqrt{\pi}\sigma e^{- (\Omega - z)^2\sigma^2} \right) \left( - \pi \mathrm{sgn}(z) \frac{\sin(( l_A n - l_B m ) z}{l_A n - l_B m } \right) \\
    &= \frac{\sigma}{4\pi} \sum_{l_A n\neq l_B m} \frac{  e^{- \frac{(l_A n - l_B m )^2}{4\sigma^2}}}{(l_A n - l_B m )} \mathrm{Im} \left[ e^{i (l_A n - l_B m ) \Omega} \mathrm{erf} \left( \frac{i(l_A n - l_B m )}{2\sigma} + \sigma \Omega \right) \right] 
\end{align}
The full expression is thus 
\begin{align}
    L_{AB} &= \frac{1}{\sum_n \eta^{2n}} \sum_{l_An = l_Bm} P_M + \frac{\sigma}{4\sqrt{\pi} \sum_n \eta^{2n}} \sum_{l_An \neq l_Bm} \frac{e^{- \frac{(l_A n - l_Bm)^2}{4\sigma^2}}}{l_An-l_Bm} \mathrm{Im} \bigg[ e^{i(l_An-l_Bm)\Omega} \mathrm{erf} \left[ \frac{i(l_A n - l_B m )}{2\sigma}  + \sigma\Omega \right] \bigg] \non \\
    & - \frac{\sigma}{2\sqrt{\pi}\sum_n \eta^{2n}} \sum_{l_An>l_Bm} \frac{e^{-\frac{(l_An-l_Bm)^2}{4\sigma^2}}}{l_An-l_Bm} \sin(\Omega(l_An-l_Bm)) 
\end{align}
as stated in the main text. 

% \textbf{Plots of the coefficient function}
% \begin{figure}[h]
%     \centering
%     \includegraphics[width=.5\linewidth]{coefficientLA12345Josh.png}
%     \caption{Plot of $\mathsf{ncoeff}f(0)$ as a function of $l_A/l_B>1$}
    
%     \label{fig:my_label}
%     \includegraphics[width=.5\textwidth]{coefficientLA.0-1-Josh.png}
%     \caption{Plot of $\mathsf{ncoeff}f(0)$ for $l_A/l_B\leq 1$. }
%     \label{fig:my_label}
% \end{figure}

\end{widetext}

\bibliography{main}

\begin{thebibliography}{38}%
\makeatletter
\providecommand \@ifxundefined [1]{%
 \@ifx{#1\undefined}
}%
\providecommand \@ifnum [1]{%
 \ifnum #1\expandafter \@firstoftwo
 \else \expandafter \@secondoftwo
 \fi
}%
\providecommand \@ifx [1]{%
 \ifx #1\expandafter \@firstoftwo
 \else \expandafter \@secondoftwo
 \fi
}%
\providecommand \natexlab [1]{#1}%
\providecommand \enquote  [1]{``#1''}%
\providecommand \bibnamefont  [1]{#1}%
\providecommand \bibfnamefont [1]{#1}%
\providecommand \citenamefont [1]{#1}%
\providecommand \href@noop [0]{\@secondoftwo}%
\providecommand \href [0]{\begingroup \@sanitize@url \@href}%
\providecommand \@href[1]{\@@startlink{#1}\@@href}%
\providecommand \@@href[1]{\endgroup#1\@@endlink}%
\providecommand \@sanitize@url [0]{\catcode `\\12\catcode `\$12\catcode
  `\&12\catcode `\#12\catcode `\^12\catcode `\_12\catcode `\%12\relax}%
\providecommand \@@startlink[1]{}%
\providecommand \@@endlink[0]{}%
\providecommand \url  [0]{\begingroup\@sanitize@url \@url }%
\providecommand \@url [1]{\endgroup\@href {#1}{\urlprefix }}%
\providecommand \urlprefix  [0]{URL }%
\providecommand \Eprint [0]{\href }%
\providecommand \doibase [0]{https://doi.org/}%
\providecommand \selectlanguage [0]{\@gobble}%
\providecommand \bibinfo  [0]{\@secondoftwo}%
\providecommand \bibfield  [0]{\@secondoftwo}%
\providecommand \translation [1]{[#1]}%
\providecommand \BibitemOpen [0]{}%
\providecommand \bibitemStop [0]{}%
\providecommand \bibitemNoStop [0]{.\EOS\space}%
\providecommand \EOS [0]{\spacefactor3000\relax}%
\providecommand \BibitemShut  [1]{\csname bibitem#1\endcsname}%
\let\auto@bib@innerbib\@empty
%</preamble>
\bibitem [{\citenamefont {Gubser}\ \emph {et~al.}(1998)\citenamefont {Gubser},
  \citenamefont {Klebanov},\ and\ \citenamefont {Polyakov}}]{gubser1998gauge}%
  \BibitemOpen
  \bibfield  {author} {\bibinfo {author} {\bibfnamefont {S.}~\bibnamefont
  {Gubser}}, \bibinfo {author} {\bibfnamefont {I.~R.}\ \bibnamefont
  {Klebanov}},\ and\ \bibinfo {author} {\bibfnamefont {A.~M.}\ \bibnamefont
  {Polyakov}},\ }\bibfield  {title} {\bibinfo {title} {{Gauge theory
  correlators from noncritical string theory}},\ }\href
  {https://doi.org/10.1016/S0370-2693(98)00377-3} {\bibfield  {journal}
  {\bibinfo  {journal} {Phys. Lett. B}\ }\textbf {\bibinfo {volume} {428}},\
  \bibinfo {pages} {105} (\bibinfo {year} {1998})},\ \Eprint
  {https://arxiv.org/abs/hep-th/9802109} {arXiv:hep-th/9802109} \BibitemShut
  {NoStop}%
\bibitem [{\citenamefont {Seiberg}\ and\ \citenamefont
  {Witten}(1999)}]{seiberg1999string}%
  \BibitemOpen
  \bibfield  {author} {\bibinfo {author} {\bibfnamefont {N.}~\bibnamefont
  {Seiberg}}\ and\ \bibinfo {author} {\bibfnamefont {E.}~\bibnamefont
  {Witten}},\ }\bibfield  {title} {\bibinfo {title} {String theory and
  noncommutative geometry},\ }\href
  {https://doi.org/10.1088/1126-6708/1999/09/032} {\bibfield  {journal}
  {\bibinfo  {journal} {Journal of High Energy Physics}\ }\textbf {\bibinfo
  {volume} {1999}},\ \bibinfo {pages} {032} (\bibinfo {year}
  {1999})}\BibitemShut {NoStop}%
\bibitem [{\citenamefont {Polchinski}(1998)}]{polchinski1998string}%
  \BibitemOpen
  \bibfield  {author} {\bibinfo {author} {\bibfnamefont {J.}~\bibnamefont
  {Polchinski}},\ }\href@noop {} {\emph {\bibinfo {title} {String theory:
  Volume 2, superstring theory and beyond}}}\ (\bibinfo  {publisher} {Cambridge
  university press},\ \bibinfo {year} {1998})\BibitemShut {NoStop}%
\bibitem [{\citenamefont {Witten}(1995)}]{witten1995string}%
  \BibitemOpen
  \bibfield  {author} {\bibinfo {author} {\bibfnamefont {E.}~\bibnamefont
  {Witten}},\ }\bibfield  {title} {\bibinfo {title} {{String theory dynamics in
  various dimensions}},\ }\href {https://doi.org/10.1016/0550-3213(95)00158-O}
  {\bibfield  {journal} {\bibinfo  {journal} {Nucl. Phys. B}\ }\textbf
  {\bibinfo {volume} {443}},\ \bibinfo {pages} {85} (\bibinfo {year}
  {1995})}\BibitemShut {NoStop}%
\bibitem [{\citenamefont {Rovelli}(1998)}]{rovelli2008loop}%
  \BibitemOpen
  \bibfield  {author} {\bibinfo {author} {\bibfnamefont {C.}~\bibnamefont
  {Rovelli}},\ }\bibfield  {title} {\bibinfo {title} {{Loop quantum gravity}},\
  }\href {https://doi.org/10.12942/lrr-1998-1} {\bibfield  {journal} {\bibinfo
  {journal} {Living Rev. Rel.}\ }\textbf {\bibinfo {volume} {1}},\ \bibinfo
  {pages} {1} (\bibinfo {year} {1998})}\BibitemShut {NoStop}%
\bibitem [{\citenamefont {Rovelli}\ and\ \citenamefont
  {Vidotto}(2014)}]{rovelli2014covariant}%
  \BibitemOpen
  \bibfield  {author} {\bibinfo {author} {\bibfnamefont {C.}~\bibnamefont
  {Rovelli}}\ and\ \bibinfo {author} {\bibfnamefont {F.}~\bibnamefont
  {Vidotto}},\ }\href@noop {} {\emph {\bibinfo {title} {Covariant loop quantum
  gravity: an elementary introduction to quantum gravity and spinfoam
  theory}}}\ (\bibinfo  {publisher} {Cambridge University Press},\ \bibinfo
  {year} {2014})\BibitemShut {NoStop}%
\bibitem [{\citenamefont {Thiemann}(2003)}]{thiemann2003lectures}%
  \BibitemOpen
  \bibfield  {author} {\bibinfo {author} {\bibfnamefont {T.}~\bibnamefont
  {Thiemann}},\ }\bibfield  {title} {\bibinfo {title} {Lectures on loop quantum
  gravity},\ }in\ \href@noop {} {\emph {\bibinfo {booktitle} {Quantum
  gravity}}}\ (\bibinfo  {publisher} {Springer},\ \bibinfo {year} {2003})\ pp.\
  \bibinfo {pages} {41--135}\BibitemShut {NoStop}%
\bibitem [{\citenamefont {Smith}\ and\ \citenamefont
  {Ahmadi}(2020)}]{smith2020quantum}%
  \BibitemOpen
  \bibfield  {author} {\bibinfo {author} {\bibfnamefont {A.~R.}\ \bibnamefont
  {Smith}}\ and\ \bibinfo {author} {\bibfnamefont {M.}~\bibnamefont {Ahmadi}},\
  }\bibfield  {title} {\bibinfo {title} {Quantum clocks observe classical and
  quantum time dilation},\ }\href@noop {} {\bibfield  {journal} {\bibinfo
  {journal} {Nature communications}\ }\textbf {\bibinfo {volume} {11}},\
  \bibinfo {pages} {1} (\bibinfo {year} {2020})}\BibitemShut {NoStop}%
\bibitem [{\citenamefont {Kempf}(2021)}]{10.3389/fphy.2021.655857}%
  \BibitemOpen
  \bibfield  {author} {\bibinfo {author} {\bibfnamefont {A.}~\bibnamefont
  {Kempf}},\ }\bibfield  {title} {\bibinfo {title} {Replacing the notion of
  spacetime distance by the notion of correlation},\ }\bibfield  {journal}
  {\bibinfo  {journal} {Frontiers in Physics}\ }\textbf {\bibinfo {volume}
  {9}},\ \href {https://doi.org/10.3389/fphy.2021.655857}
  {10.3389/fphy.2021.655857} (\bibinfo {year} {2021})\BibitemShut {NoStop}%
\bibitem [{\citenamefont {Zych}\ \emph {et~al.}(2019)\citenamefont {Zych},
  \citenamefont {Costa}, \citenamefont {Pikovski},\ and\ \citenamefont
  {Brukner}}]{zych2019bell}%
  \BibitemOpen
  \bibfield  {author} {\bibinfo {author} {\bibfnamefont {M.}~\bibnamefont
  {Zych}}, \bibinfo {author} {\bibfnamefont {F.}~\bibnamefont {Costa}},
  \bibinfo {author} {\bibfnamefont {I.}~\bibnamefont {Pikovski}},\ and\
  \bibinfo {author} {\bibfnamefont {v.~C.}\ \bibnamefont {Brukner}},\
  }\bibfield  {title} {\bibinfo {title} {{Bell's theorem for temporal order}},\
  }\href {https://doi.org/10.1038/s41467-019-11579-x} {\bibfield  {journal}
  {\bibinfo  {journal} {Nature Commun.}\ }\textbf {\bibinfo {volume} {10}},\
  \bibinfo {pages} {3772} (\bibinfo {year} {2019})},\ \Eprint
  {https://arxiv.org/abs/1708.00248} {arXiv:1708.00248 [quant-ph]} \BibitemShut
  {NoStop}%
\bibitem [{\citenamefont {Christodoulou}\ and\ \citenamefont
  {Rovelli}(2019)}]{christodoulou2019possibility}%
  \BibitemOpen
  \bibfield  {author} {\bibinfo {author} {\bibfnamefont {M.}~\bibnamefont
  {Christodoulou}}\ and\ \bibinfo {author} {\bibfnamefont {C.}~\bibnamefont
  {Rovelli}},\ }\bibfield  {title} {\bibinfo {title} {On the possibility of
  laboratory evidence for quantum superposition of geometries},\ }\href@noop {}
  {\bibfield  {journal} {\bibinfo  {journal} {Physics Letters B}\ }\textbf
  {\bibinfo {volume} {792}},\ \bibinfo {pages} {64} (\bibinfo {year}
  {2019})}\BibitemShut {NoStop}%
\bibitem [{\citenamefont {Belenchia}\ \emph {et~al.}(2018)\citenamefont
  {Belenchia}, \citenamefont {Wald}, \citenamefont {Giacomini}, \citenamefont
  {Castro-Ruiz}, \citenamefont {Brukner},\ and\ \citenamefont
  {Aspelmeyer}}]{belenchiaPhysRevD.98.126009}%
  \BibitemOpen
  \bibfield  {author} {\bibinfo {author} {\bibfnamefont {A.}~\bibnamefont
  {Belenchia}}, \bibinfo {author} {\bibfnamefont {R.~M.}\ \bibnamefont {Wald}},
  \bibinfo {author} {\bibfnamefont {F.}~\bibnamefont {Giacomini}}, \bibinfo
  {author} {\bibfnamefont {E.}~\bibnamefont {Castro-Ruiz}}, \bibinfo {author}
  {\bibfnamefont {i.~c.~v.}\ \bibnamefont {Brukner}},\ and\ \bibinfo {author}
  {\bibfnamefont {M.}~\bibnamefont {Aspelmeyer}},\ }\bibfield  {title}
  {\bibinfo {title} {Quantum superposition of massive objects and the
  quantization of gravity},\ }\href
  {https://doi.org/10.1103/PhysRevD.98.126009} {\bibfield  {journal} {\bibinfo
  {journal} {Phys. Rev. D}\ }\textbf {\bibinfo {volume} {98}},\ \bibinfo
  {pages} {126009} (\bibinfo {year} {2018})}\BibitemShut {NoStop}%
\bibitem [{\citenamefont {Giacomini}(2021)}]{Giacomini2021spacetimequantum}%
  \BibitemOpen
  \bibfield  {author} {\bibinfo {author} {\bibfnamefont {F.}~\bibnamefont
  {Giacomini}},\ }\bibfield  {title} {\bibinfo {title} {Spacetime {Q}uantum
  {R}eference {F}rames and superpositions of proper times},\ }\href
  {https://doi.org/10.22331/q-2021-07-22-508} {\bibfield  {journal} {\bibinfo
  {journal} {{Quantum}}\ }\textbf {\bibinfo {volume} {5}},\ \bibinfo {pages}
  {508} (\bibinfo {year} {2021})}\BibitemShut {NoStop}%
\bibitem [{\citenamefont {Giacomini}\ and\ \citenamefont
  {Brukner}(2022)}]{giacomini2022quantum}%
  \BibitemOpen
  \bibfield  {author} {\bibinfo {author} {\bibfnamefont {F.}~\bibnamefont
  {Giacomini}}\ and\ \bibinfo {author} {\bibfnamefont {{\v{C}}.}~\bibnamefont
  {Brukner}},\ }\bibfield  {title} {\bibinfo {title} {Quantum superposition of
  spacetimes obeys einstein's equivalence principle},\ }\href@noop {}
  {\bibfield  {journal} {\bibinfo  {journal} {AVS Quantum Science}\ }\textbf
  {\bibinfo {volume} {4}},\ \bibinfo {pages} {015601} (\bibinfo {year}
  {2022})}\BibitemShut {NoStop}%
\bibitem [{\citenamefont {Foo}\ \emph {et~al.}(2021{\natexlab{a}})\citenamefont
  {Foo}, \citenamefont {Mann},\ and\ \citenamefont {Zych}}]{Foo_2021}%
  \BibitemOpen
  \bibfield  {author} {\bibinfo {author} {\bibfnamefont {J.}~\bibnamefont
  {Foo}}, \bibinfo {author} {\bibfnamefont {R.~B.}\ \bibnamefont {Mann}},\ and\
  \bibinfo {author} {\bibfnamefont {M.}~\bibnamefont {Zych}},\ }\bibfield
  {title} {\bibinfo {title} {Schrödinger's cat for de sitter spacetime},\
  }\href {https://doi.org/10.1088/1361-6382/abf1c4} {\bibfield  {journal}
  {\bibinfo  {journal} {Classical and Quantum Gravity}\ }\textbf {\bibinfo
  {volume} {38}},\ \bibinfo {pages} {115010} (\bibinfo {year}
  {2021}{\natexlab{a}})}\BibitemShut {NoStop}%
\bibitem [{\citenamefont {Foo}\ \emph {et~al.}(2021{\natexlab{b}})\citenamefont
  {Foo}, \citenamefont {Arabaci}, \citenamefont {Zych},\ and\ \citenamefont
  {Mann}}]{foo_2022}%
  \BibitemOpen
  \bibfield  {author} {\bibinfo {author} {\bibfnamefont {J.}~\bibnamefont
  {Foo}}, \bibinfo {author} {\bibfnamefont {C.~S.}\ \bibnamefont {Arabaci}},
  \bibinfo {author} {\bibfnamefont {M.}~\bibnamefont {Zych}},\ and\ \bibinfo
  {author} {\bibfnamefont {R.~B.}\ \bibnamefont {Mann}},\ }\href
  {https://doi.org/10.48550/ARXIV.2111.13315} {\bibinfo {title} {Quantum
  signatures of black hole mass superpositions}} (\bibinfo {year}
  {2021}{\natexlab{b}})\BibitemShut {NoStop}%
\bibitem [{\citenamefont {Bekenstein}(2020)}]{bekenstein2020quantum}%
  \BibitemOpen
  \bibfield  {author} {\bibinfo {author} {\bibfnamefont {J.~D.}\ \bibnamefont
  {Bekenstein}},\ }\bibfield  {title} {\bibinfo {title} {The quantum mass
  spectrum of the kerr black hole},\ }in\ \href@noop {} {\emph {\bibinfo
  {booktitle} {JACOB BEKENSTEIN: The Conservative Revolutionary}}}\ (\bibinfo
  {publisher} {World Scientific},\ \bibinfo {year} {2020})\ pp.\ \bibinfo
  {pages} {331--334}\BibitemShut {NoStop}%
\bibitem [{\citenamefont {Bekenstein}(1973)}]{bekensteinPhysRevD.7.2333}%
  \BibitemOpen
  \bibfield  {author} {\bibinfo {author} {\bibfnamefont {J.~D.}\ \bibnamefont
  {Bekenstein}},\ }\bibfield  {title} {\bibinfo {title} {Black holes and
  entropy},\ }\href {https://doi.org/10.1103/PhysRevD.7.2333} {\bibfield
  {journal} {\bibinfo  {journal} {Phys. Rev. D}\ }\textbf {\bibinfo {volume}
  {7}},\ \bibinfo {pages} {2333} (\bibinfo {year} {1973})}\BibitemShut
  {NoStop}%
\bibitem [{\citenamefont {Bowen}\ and\ \citenamefont
  {Milburn}(2015)}]{bowen2015quantum}%
  \BibitemOpen
  \bibfield  {author} {\bibinfo {author} {\bibfnamefont {W.~P.}\ \bibnamefont
  {Bowen}}\ and\ \bibinfo {author} {\bibfnamefont {G.~J.}\ \bibnamefont
  {Milburn}},\ }\href@noop {} {\emph {\bibinfo {title} {Quantum
  optomechanics}}}\ (\bibinfo  {publisher} {CRC press},\ \bibinfo {year}
  {2015})\BibitemShut {NoStop}%
\bibitem [{\citenamefont {Barcel{\'o}}\ \emph {et~al.}(2021)\citenamefont
  {Barcel{\'o}}, \citenamefont {Garay},\ and\ \citenamefont
  {Garc{\'\i}a-Moreno}}]{barcelo2021superposing}%
  \BibitemOpen
  \bibfield  {author} {\bibinfo {author} {\bibfnamefont {C.}~\bibnamefont
  {Barcel{\'o}}}, \bibinfo {author} {\bibfnamefont {L.~J.}\ \bibnamefont
  {Garay}},\ and\ \bibinfo {author} {\bibfnamefont {G.}~\bibnamefont
  {Garc{\'\i}a-Moreno}},\ }\bibfield  {title} {\bibinfo {title} {Superposing
  spacetimes: lessons from analogue gravity},\ }\href@noop {} {\bibfield
  {journal} {\bibinfo  {journal} {arXiv preprint arXiv:2104.15078}\ } (\bibinfo
  {year} {2021})}\BibitemShut {NoStop}%
\bibitem [{\citenamefont {Mart\'{\i}n-Mart\'{\i}nez}\ \emph
  {et~al.}(2016{\natexlab{a}})\citenamefont {Mart\'{\i}n-Mart\'{\i}nez},
  \citenamefont {Smith},\ and\ \citenamefont
  {Terno}}]{martinmartinezPhysRevD.93.044001}%
  \BibitemOpen
  \bibfield  {author} {\bibinfo {author} {\bibfnamefont {E.}~\bibnamefont
  {Mart\'{\i}n-Mart\'{\i}nez}}, \bibinfo {author} {\bibfnamefont {A.~R.~H.}\
  \bibnamefont {Smith}},\ and\ \bibinfo {author} {\bibfnamefont {D.~R.}\
  \bibnamefont {Terno}},\ }\bibfield  {title} {\bibinfo {title} {Spacetime
  structure and vacuum entanglement},\ }\href
  {https://doi.org/10.1103/PhysRevD.93.044001} {\bibfield  {journal} {\bibinfo
  {journal} {Phys. Rev. D}\ }\textbf {\bibinfo {volume} {93}},\ \bibinfo
  {pages} {044001} (\bibinfo {year} {2016}{\natexlab{a}})}\BibitemShut
  {NoStop}%
\bibitem [{\citenamefont {Banach}\ and\ \citenamefont
  {Dowker}(1979)}]{Banach:1979iy}%
  \BibitemOpen
  \bibfield  {author} {\bibinfo {author} {\bibfnamefont {R.}~\bibnamefont
  {Banach}}\ and\ \bibinfo {author} {\bibfnamefont {J.~S.}\ \bibnamefont
  {Dowker}},\ }\bibfield  {title} {\bibinfo {title} {{The Vacuum Stress Tensor
  for Automorphic Fields on Some Flat Space-times}},\ }\href
  {https://doi.org/10.1088/0305-4470/12/12/032} {\bibfield  {journal} {\bibinfo
   {journal} {J. Phys. A}\ }\textbf {\bibinfo {volume} {12}},\ \bibinfo {pages}
  {2545} (\bibinfo {year} {1979})}\BibitemShut {NoStop}%
\bibitem [{\citenamefont {Banach}(1980)}]{Banach_1980}%
  \BibitemOpen
  \bibfield  {author} {\bibinfo {author} {\bibfnamefont {R.}~\bibnamefont
  {Banach}},\ }\bibfield  {title} {\bibinfo {title} {The quantum theory of free
  automorphic fields},\ }\href {https://doi.org/10.1088/0305-4470/13/6/039}
  {\bibfield  {journal} {\bibinfo  {journal} {Journal of Physics A:
  Mathematical and General}\ }\textbf {\bibinfo {volume} {13}},\ \bibinfo
  {pages} {2179} (\bibinfo {year} {1980})}\BibitemShut {NoStop}%
\bibitem [{\citenamefont {Langlois}(2006)}]{langlois2006causal}%
  \BibitemOpen
  \bibfield  {author} {\bibinfo {author} {\bibfnamefont {P.}~\bibnamefont
  {Langlois}},\ }\bibfield  {title} {\bibinfo {title} {Causal particle
  detectors and topology},\ }\href@noop {} {\bibfield  {journal} {\bibinfo
  {journal} {Annals of Physics}\ }\textbf {\bibinfo {volume} {321}},\ \bibinfo
  {pages} {2027} (\bibinfo {year} {2006})}\BibitemShut {NoStop}%
\bibitem [{\citenamefont {Kabel}\ \emph {et~al.}(2022)\citenamefont {Kabel},
  \citenamefont {de~la Hamette}, \citenamefont {Castro-Ruiz},\ and\
  \citenamefont {Brukner}}]{kabel}%
  \BibitemOpen
  \bibfield  {author} {\bibinfo {author} {\bibfnamefont {V.}~\bibnamefont
  {Kabel}}, \bibinfo {author} {\bibfnamefont {A.-C.}\ \bibnamefont {de~la
  Hamette}}, \bibinfo {author} {\bibfnamefont {E.}~\bibnamefont
  {Castro-Ruiz}},\ and\ \bibinfo {author} {\bibfnamefont {C.}~\bibnamefont
  {Brukner}},\ }\href {https://doi.org/10.48550/ARXIV.2207.00021} {\bibinfo
  {title} {Quantum conformal symmetries for spacetimes in superposition}}
  (\bibinfo {year} {2022})\BibitemShut {NoStop}%
\bibitem [{\citenamefont {Foo}\ \emph {et~al.}(2020)\citenamefont {Foo},
  \citenamefont {Onoe},\ and\ \citenamefont
  {Zych}}]{fooudw1PhysRevD.102.085013}%
  \BibitemOpen
  \bibfield  {author} {\bibinfo {author} {\bibfnamefont {J.}~\bibnamefont
  {Foo}}, \bibinfo {author} {\bibfnamefont {S.}~\bibnamefont {Onoe}},\ and\
  \bibinfo {author} {\bibfnamefont {M.}~\bibnamefont {Zych}},\ }\bibfield
  {title} {\bibinfo {title} {Unruh-dewitt detectors in quantum superpositions
  of trajectories},\ }\href {https://doi.org/10.1103/PhysRevD.102.085013}
  {\bibfield  {journal} {\bibinfo  {journal} {Phys. Rev. D}\ }\textbf {\bibinfo
  {volume} {102}},\ \bibinfo {pages} {085013} (\bibinfo {year}
  {2020})}\BibitemShut {NoStop}%
\bibitem [{\citenamefont {Foo}\ \emph {et~al.}(2021{\natexlab{c}})\citenamefont
  {Foo}, \citenamefont {Onoe}, \citenamefont {Mann},\ and\ \citenamefont
  {Zych}}]{fooudw2PhysRevResearch.3.043056}%
  \BibitemOpen
  \bibfield  {author} {\bibinfo {author} {\bibfnamefont {J.}~\bibnamefont
  {Foo}}, \bibinfo {author} {\bibfnamefont {S.}~\bibnamefont {Onoe}}, \bibinfo
  {author} {\bibfnamefont {R.~B.}\ \bibnamefont {Mann}},\ and\ \bibinfo
  {author} {\bibfnamefont {M.}~\bibnamefont {Zych}},\ }\bibfield  {title}
  {\bibinfo {title} {Thermality, causality, and the quantum-controlled
  unruh--dewitt detector},\ }\href
  {https://doi.org/10.1103/PhysRevResearch.3.043056} {\bibfield  {journal}
  {\bibinfo  {journal} {Phys. Rev. Research}\ }\textbf {\bibinfo {volume}
  {3}},\ \bibinfo {pages} {043056} (\bibinfo {year}
  {2021}{\natexlab{c}})}\BibitemShut {NoStop}%
\bibitem [{\citenamefont {Foo}\ \emph {et~al.}(2021{\natexlab{d}})\citenamefont
  {Foo}, \citenamefont {Mann},\ and\ \citenamefont
  {Zych}}]{fooudw3PhysRevD.103.065013}%
  \BibitemOpen
  \bibfield  {author} {\bibinfo {author} {\bibfnamefont {J.}~\bibnamefont
  {Foo}}, \bibinfo {author} {\bibfnamefont {R.~B.}\ \bibnamefont {Mann}},\ and\
  \bibinfo {author} {\bibfnamefont {M.}~\bibnamefont {Zych}},\ }\bibfield
  {title} {\bibinfo {title} {Entanglement amplification between superposed
  detectors in flat and curved spacetimes},\ }\href
  {https://doi.org/10.1103/PhysRevD.103.065013} {\bibfield  {journal} {\bibinfo
   {journal} {Phys. Rev. D}\ }\textbf {\bibinfo {volume} {103}},\ \bibinfo
  {pages} {065013} (\bibinfo {year} {2021}{\natexlab{d}})}\BibitemShut
  {NoStop}%
\bibitem [{\citenamefont {Foo}\ \emph {et~al.}(2022)\citenamefont {Foo},
  \citenamefont {Mann},\ and\ \citenamefont
  {Zych}}]{fooudw5https://doi.org/10.48550/arxiv.2204.00384}%
  \BibitemOpen
  \bibfield  {author} {\bibinfo {author} {\bibfnamefont {J.}~\bibnamefont
  {Foo}}, \bibinfo {author} {\bibfnamefont {R.~B.}\ \bibnamefont {Mann}},\ and\
  \bibinfo {author} {\bibfnamefont {M.}~\bibnamefont {Zych}},\ }\href
  {https://doi.org/10.48550/ARXIV.2204.00384} {\bibinfo {title} {Schrödinger's
  black hole cat}} (\bibinfo {year} {2022})\BibitemShut {NoStop}%
\bibitem [{\citenamefont {Howl}\ \emph {et~al.}(2022)\citenamefont {Howl},
  \citenamefont {Akil}, \citenamefont {Kristjánsson}, \citenamefont {Zhao},\
  and\ \citenamefont
  {Chiribella}}]{howlhttps://doi.org/10.48550/arxiv.2203.05861}%
  \BibitemOpen
  \bibfield  {author} {\bibinfo {author} {\bibfnamefont {R.}~\bibnamefont
  {Howl}}, \bibinfo {author} {\bibfnamefont {A.}~\bibnamefont {Akil}}, \bibinfo
  {author} {\bibfnamefont {H.}~\bibnamefont {Kristjánsson}}, \bibinfo {author}
  {\bibfnamefont {X.}~\bibnamefont {Zhao}},\ and\ \bibinfo {author}
  {\bibfnamefont {G.}~\bibnamefont {Chiribella}},\ }\href
  {https://doi.org/10.48550/ARXIV.2203.05861} {\bibinfo {title} {Quantum
  gravity as a communication resource}} (\bibinfo {year} {2022})\BibitemShut
  {NoStop}%
\bibitem [{\citenamefont {Whaley}\ and\ \citenamefont
  {Light}(1984)}]{whaleyPhysRevA.29.1188}%
  \BibitemOpen
  \bibfield  {author} {\bibinfo {author} {\bibfnamefont {K.~B.}\ \bibnamefont
  {Whaley}}\ and\ \bibinfo {author} {\bibfnamefont {J.~C.}\ \bibnamefont
  {Light}},\ }\bibfield  {title} {\bibinfo {title} {Rotating-frame
  transformations: A new approximation for multiphoton absorption and
  dissociation in laser fields},\ }\href
  {https://doi.org/10.1103/PhysRevA.29.1188} {\bibfield  {journal} {\bibinfo
  {journal} {Phys. Rev. A}\ }\textbf {\bibinfo {volume} {29}},\ \bibinfo
  {pages} {1188} (\bibinfo {year} {1984})}\BibitemShut {NoStop}%
\bibitem [{\citenamefont {Steeg}\ and\ \citenamefont
  {Menicucci}(2009)}]{menicucciPhysRevD.79.044027}%
  \BibitemOpen
  \bibfield  {author} {\bibinfo {author} {\bibfnamefont {G.~V.}\ \bibnamefont
  {Steeg}}\ and\ \bibinfo {author} {\bibfnamefont {N.~C.}\ \bibnamefont
  {Menicucci}},\ }\bibfield  {title} {\bibinfo {title} {Entangling power of an
  expanding universe},\ }\href {https://doi.org/10.1103/PhysRevD.79.044027}
  {\bibfield  {journal} {\bibinfo  {journal} {Phys. Rev. D}\ }\textbf {\bibinfo
  {volume} {79}},\ \bibinfo {pages} {044027} (\bibinfo {year}
  {2009})}\BibitemShut {NoStop}%
\bibitem [{\citenamefont {Mart\'{\i}n-Mart\'{\i}nez}\ \emph
  {et~al.}(2016{\natexlab{b}})\citenamefont {Mart\'{\i}n-Mart\'{\i}nez},
  \citenamefont {Smith},\ and\ \citenamefont {Terno}}]{PhysRevD.93.044001}%
  \BibitemOpen
  \bibfield  {author} {\bibinfo {author} {\bibfnamefont {E.}~\bibnamefont
  {Mart\'{\i}n-Mart\'{\i}nez}}, \bibinfo {author} {\bibfnamefont {A.~R.~H.}\
  \bibnamefont {Smith}},\ and\ \bibinfo {author} {\bibfnamefont {D.~R.}\
  \bibnamefont {Terno}},\ }\bibfield  {title} {\bibinfo {title} {Spacetime
  structure and vacuum entanglement},\ }\href
  {https://doi.org/10.1103/PhysRevD.93.044001} {\bibfield  {journal} {\bibinfo
  {journal} {Phys. Rev. D}\ }\textbf {\bibinfo {volume} {93}},\ \bibinfo
  {pages} {044001} (\bibinfo {year} {2016}{\natexlab{b}})}\BibitemShut
  {NoStop}%
\bibitem [{\citenamefont {Tjoa}\ and\ \citenamefont
  {Mart\'{\i}n-Mart\'{\i}nez}(2020)}]{tjoaPhysRevD.101.125020}%
  \BibitemOpen
  \bibfield  {author} {\bibinfo {author} {\bibfnamefont {E.}~\bibnamefont
  {Tjoa}}\ and\ \bibinfo {author} {\bibfnamefont {E.}~\bibnamefont
  {Mart\'{\i}n-Mart\'{\i}nez}},\ }\bibfield  {title} {\bibinfo {title} {Vacuum
  entanglement harvesting with a zero mode},\ }\href
  {https://doi.org/10.1103/PhysRevD.101.125020} {\bibfield  {journal} {\bibinfo
   {journal} {Phys. Rev. D}\ }\textbf {\bibinfo {volume} {101}},\ \bibinfo
  {pages} {125020} (\bibinfo {year} {2020})}\BibitemShut {NoStop}%
\bibitem [{\citenamefont {Tjoa}\ and\ \citenamefont
  {Mart\'{\i}n-Mart\'{\i}nez}(2019)}]{tjoaPhysRevD.99.065005}%
  \BibitemOpen
  \bibfield  {author} {\bibinfo {author} {\bibfnamefont {E.}~\bibnamefont
  {Tjoa}}\ and\ \bibinfo {author} {\bibfnamefont {E.}~\bibnamefont
  {Mart\'{\i}n-Mart\'{\i}nez}},\ }\bibfield  {title} {\bibinfo {title} {Zero
  mode suppression of superluminal signals in light-matter interactions},\
  }\href {https://doi.org/10.1103/PhysRevD.99.065005} {\bibfield  {journal}
  {\bibinfo  {journal} {Phys. Rev. D}\ }\textbf {\bibinfo {volume} {99}},\
  \bibinfo {pages} {065005} (\bibinfo {year} {2019})}\BibitemShut {NoStop}%
\bibitem [{\citenamefont {Brown}\ \emph {et~al.}(2015)\citenamefont {Brown},
  \citenamefont {del Rey}, \citenamefont {Westman}, \citenamefont {Le\'on},\
  and\ \citenamefont {Dragan}}]{brownPhysRevD.91.016005}%
  \BibitemOpen
  \bibfield  {author} {\bibinfo {author} {\bibfnamefont {E.~G.}\ \bibnamefont
  {Brown}}, \bibinfo {author} {\bibfnamefont {M.}~\bibnamefont {del Rey}},
  \bibinfo {author} {\bibfnamefont {H.}~\bibnamefont {Westman}}, \bibinfo
  {author} {\bibfnamefont {J.}~\bibnamefont {Le\'on}},\ and\ \bibinfo {author}
  {\bibfnamefont {A.}~\bibnamefont {Dragan}},\ }\bibfield  {title} {\bibinfo
  {title} {What does it mean for half of an empty cavity to be full?},\ }\href
  {https://doi.org/10.1103/PhysRevD.91.016005} {\bibfield  {journal} {\bibinfo
  {journal} {Phys. Rev. D}\ }\textbf {\bibinfo {volume} {91}},\ \bibinfo
  {pages} {016005} (\bibinfo {year} {2015})}\BibitemShut {NoStop}%
\bibitem [{\citenamefont {V{\'a}zquez}\ \emph {et~al.}(2014)\citenamefont
  {V{\'a}zquez}, \citenamefont {del Rey}, \citenamefont {Westman},\ and\
  \citenamefont {Le{\'o}n}}]{vazquez2014local}%
  \BibitemOpen
  \bibfield  {author} {\bibinfo {author} {\bibfnamefont {M.~R.}\ \bibnamefont
  {V{\'a}zquez}}, \bibinfo {author} {\bibfnamefont {M.}~\bibnamefont {del
  Rey}}, \bibinfo {author} {\bibfnamefont {H.}~\bibnamefont {Westman}},\ and\
  \bibinfo {author} {\bibfnamefont {J.}~\bibnamefont {Le{\'o}n}},\ }\bibfield
  {title} {\bibinfo {title} {Local quanta, unitary inequivalence, and vacuum
  entanglement},\ }\href@noop {} {\bibfield  {journal} {\bibinfo  {journal}
  {Annals of Physics}\ }\textbf {\bibinfo {volume} {351}},\ \bibinfo {pages}
  {112} (\bibinfo {year} {2014})}\BibitemShut {NoStop}%
\bibitem [{\citenamefont {Birrell}\ \emph {et~al.}(1984)\citenamefont
  {Birrell}, \citenamefont {Birrell},\ and\ \citenamefont
  {Davies}}]{birrell1984quantum}%
  \BibitemOpen
  \bibfield  {author} {\bibinfo {author} {\bibfnamefont {N.~D.}\ \bibnamefont
  {Birrell}}, \bibinfo {author} {\bibfnamefont {N.~D.}\ \bibnamefont
  {Birrell}},\ and\ \bibinfo {author} {\bibfnamefont {P.}~\bibnamefont
  {Davies}},\ }\href@noop {} {\emph {\bibinfo {title} {Quantum fields in curved
  space}}}\ (\bibinfo  {publisher} {Cambridge university press},\ \bibinfo
  {year} {1984})\BibitemShut {NoStop}%
\end{thebibliography}%

\end{document}